# Direct generation of time-energy-entangled W triphotons in atomic vapor

Kangkang Li,[1] Jianming Wen,[2]* Yin Cai,[1]* Saeid Vashahri Ghamsari[2], Changbiao Li[1], Feng Li[1], Zhaoyang Zhang[1], Yanpeng Zhang[1*], and Min Xiao[3,4]

[1]Key Laboratory for Physical Electronics and Devices of the Ministry of Education & Shaanxi Key Lab of Information Photonic Technique, Xi'an Jiaotong University, Xi'an 710049, China.
[2]Department of Physics, Kennesaw State University, Marietta, Georgia 30060, USA.
[3]National Laboratory of Solid State Microstructures, College of Engineering and Applied Sciences and School of Physics, Nanjing University, Nanjing 210093, China.
[4]Department of Physics, University of Arkansas, Fayetteville, Arkansas 72701, USA.
*Email: jianming.wen@kennesaw.edu; caiyin@xjtu.edu.cn; ypzhang@mail.xjtu.edu.cn.

## Abstract
Entangled multiphoton sources are essential for both fundamental tests of quantum foundations and building blocks of contemporary optical quantum technologies. While efforts over the past three decades have focused on creating multiphoton entanglement through multiplexing existing biphoton sources with linear optics and postselections, our work presents a groundbreaking approach. For the first time, we observe genuine continuous-mode time-energy-entangled W-class triphotons with an unprecedented production rate directly generated through spontaneous six-wave mixing (SSWM) in a four-level triple-Λ atomic vapor cell. Utilizing electromagnetically induced transparency and coherence control, our SSWM scheme allows versatile narrowband triphoton generation with advantageous properties, including long temporal coherence and controllable waveforms. This advancement is ideal for applications like long-distance quantum communications and information processing, bridging single photons and neutral atoms. Most significantly, our work establishes a reliable and efficient genuine triphoton source, facilitating accessible research on multiphoton entanglement.

## Introduction
Generating entangled multiphoton states (*1*) is pivotal to probe quantum foundations and advance technological innovations. Comprehensive studies have already shown that multiphoton entanglement (*1*) enables a plethora of classically impossible phenomena, most of them incomprehensible with any bipartite system. Unfortunately, we hitherto have at hand only biphoton sources based upon spontaneous parametric down-conversion (SPDC) or spontaneous four-wave mixing (SFWM). This has urged tremendous efforts on developing multiphoton sources (*1*−*3*) over past thirty years. Among them, the most popular means is to multiplex existing biphoton sources with linear optics and postselections. This brings us the well-known exemplar of polarization-entangled multiphotons (*4*−*8*) by constructing imperative interferometric setups. Although postselection might be acceptable in some protocols, it is generally deleterious for most applications since the action of observing photons alters and destroys the states. To avoid postselection, the second path considers cascaded SPDCs/SFWMs (*9*−*12*) or two SPDCs/SFWMs followed by one up-conversion (*13*, *14*). In this way, polarization or time-energy entangled triphotons were reported by building sophisticated coincidence counting circuits. Despite no needs on interferometric settings, the attained states are intrinsically non-Gaussian due to unbalanced photon numbers between the primary and secondary biphoton process, thereby making these sources very noisy and inefficient. Alternatively, the third technique (*15*−*17*) suggests to coherently mix paired photons with singles attenuated from a cw laser to trigger triphoton events. Akin to the first method, this solution depends on erasing the photon distinguishability by resorting to the Hong-Ou-Mandal interference effect (*18*). Though polarization-entangled multiphotons of inequivalent classes were experimented with postselection, the low success rate and required interferometric stabilization make this proposal not so practical. As photons are always emitted in pairs in SPDC/SFWM, this attribute results in the fourth route (*19*−*23*) to make use of emission of multiple pairs by appropriately setting input pump powers. Though it seems easy to yield even-number states, yet, dominant biphotons from lower-order perturbation of the parametric process challenge detecting entangled multiphotons from higher-order perturbations. To have an acceptable fidelity, like the second way, a complicated detection system plus an interferometric setup is often inevitable in practice. What's more, this approach mainly allows to form polarization entanglement thus far. In spite of these impressive achievements, all foregoing mechanisms are difficult to offer a reliable and efficient triphoton source for research and applications. Additionally, so far there is no convincing realization of the entangled triphoton experiment in continuous modes. Driven by SPDC, one would expect that such photons could be naturally born from third-order SPDC (*24*, *25*) by converting one pump photon of higher energy into three daughter photons of low energy. The idea looks simple and straightforward, but experimentally inaccessible owing to the lack of such a nonlinear optical material. As a result, developing a reliable triphoton source is still in its infancy even up to today.

Coherent atomic media (*26*), on the other hand, exhibit a wide range of peculiar properties including giant nonlinearities, prolonged atomic coherence, strong photon-atom interaction, and slow/fast light effects. Recently, these exotic properties have been skillfully employed to construct a novel narrowband biphoton source (*27*−*30*) basing on SFWM. Specifically, giant nonlinearities promise efficient parametric conversion, long atomic coherence leads to narrowband wavepackets, and sharp optical response becomes a formidable knob for shaping photon waveforms and temporal correlations. Unlike solid state sources, one unique feature pertinent to atomic ensembles arises from the dual role played by the third-order nonlinear susceptibility $\chi^{(3)}$ in biphoton generation (*27*, *31*−*33*). That is, in addition to governing nonlinear conversion strength, the double-resonance structure in $\chi^{(3)}$ signifies the coexistence of two sets of SFWMs in light quanta radiation. Alternatively, entangled photons output from these two stochastic but coherent SFWM processes interfere and give rise to a nontrivial two-photon interference, namely, the damped Rabi oscillations. In general, their waveforms are entirely patterned by the convolution of a complex phase-mismatch function and $\chi^{(3)}$. Other than these attributes, the nonclassical correlations shared by paired photons can be additionally manipulated by exploiting various coherent control techniques including electromagnetically induced transparency (*26*) (EIT) to reshape optical responses. The interplay amongst diverse effects also enriches fundamental research and fosters technological innovations, inaccessible to other existing biphoton sources. Besides, flexible system layouts like backward detection geometry are more favorable to photon counting detection. Motivated by these advantages, here we move one step forward and report the direct generation of continuous-mode triphotons entangled in time and energy from a hot atomic vapor cell. By utilizing the process of spontaneous six-wave mixing (SSWM) (*34*, *35*), we have not only observed the striking three-photon interference but also witnessed the residual two~photon correlation by tracing one photon out, an intrinsic virtue of the W class of tripartite entanglement (*34*). By adjusting the system parameters, we have further achieved waveform-controllable triphoton generation. Together with an unprecedented production rate, our scheme has substantiated to be the first reliable platform that leverages multipartite entanglement research to an unparalleled level.

## Results

As schematic in Figs. 1A-C, we are interested in yielding narrowband W triphotons from a 7-cm long [85]Rb vapor cell with a four-level triple-Λ atomic configuration at temperature 80°C (or 115°C). The detail of the experimental setup is provided in Methods. In the presence of three counter-propagating cw laser beams (one weak pump $(E_1, \omega_1, \vec{k}_1)$ and two strong couplings $(E_2, \omega_2, \vec{k}_2)$ and $(E_3, \omega_3, \vec{k}_3)$), backward photon triplets $(E_{Sj}, \omega_{Sj}, \vec{k}_{Sj}$ with $j = 1, 2, 3)$ are emitted via Doppler-broadened SSWM at an intersection angle of $\theta \approx 4°$ to the principle $z$-axis along the phase matching direction, $\Delta\vec{k} = (\vec{k}_{S1} + \vec{k}_{S2} + \vec{k}_{S3}) - (\vec{k}_1 + \vec{k}_2 + \vec{k}_3) = 0$. As depicted in Figs. 1B and C, the three coaxial input lasers were coupled into the center of the [85]Rb vapor cell with tunable frequency detunings $\Delta_j$ and powers $P_j$; while the generated photon triplets were accordingly detected by three single-photon counting modules ($SPCM_1 - SPCM_3$) for coincidence counts after spatial and frequency filtering. Here, to avoid unwanted accidental trigger events induced by singles and dual biphotons, we placed single-band filters and narrowband etalon Fabry-Perot cavities in front of $SPCM_j$ before detection. We notice that in three-photon joint clicks, the major source of accidental coincidences stems from double pairs from two different SFWMs simultaneously present in the detection system (Supplementary Information (SI)). Since these dual pairs may have similar central frequencies and polarizations as genuine triphoton modes, they cannot be filtered away simply by polarizers and frequency filters. To exclude such double-pair false trigger events, in experiment we further introduced an additional $SPCM_d$ synchronized with $SPCM_3$ to serve as the diagnosis detector in conjunction with the rest two, $SPCM_1$ and $SPCM_2$. To ensure the atomic population to be mainly distributed in the ground level $\left|5S_{\frac{1}{2}}, F=2\right\rangle$ throughout the measurement, an additional strong optical repumping beam ($E_{op}$) was applied to the atomic transition $\left|5S_{\frac{1}{2}}, F=3\right\rangle \rightarrow \left|5P_{\frac{1}{2}}\right\rangle$ in alignment with $E_2$ but without spatial overlap. With these preparations, we carefully adjust the system parameters, especially $P_j$ and $\Delta_j$ of each input field $E_j$, to promote the SSWM occurrence.

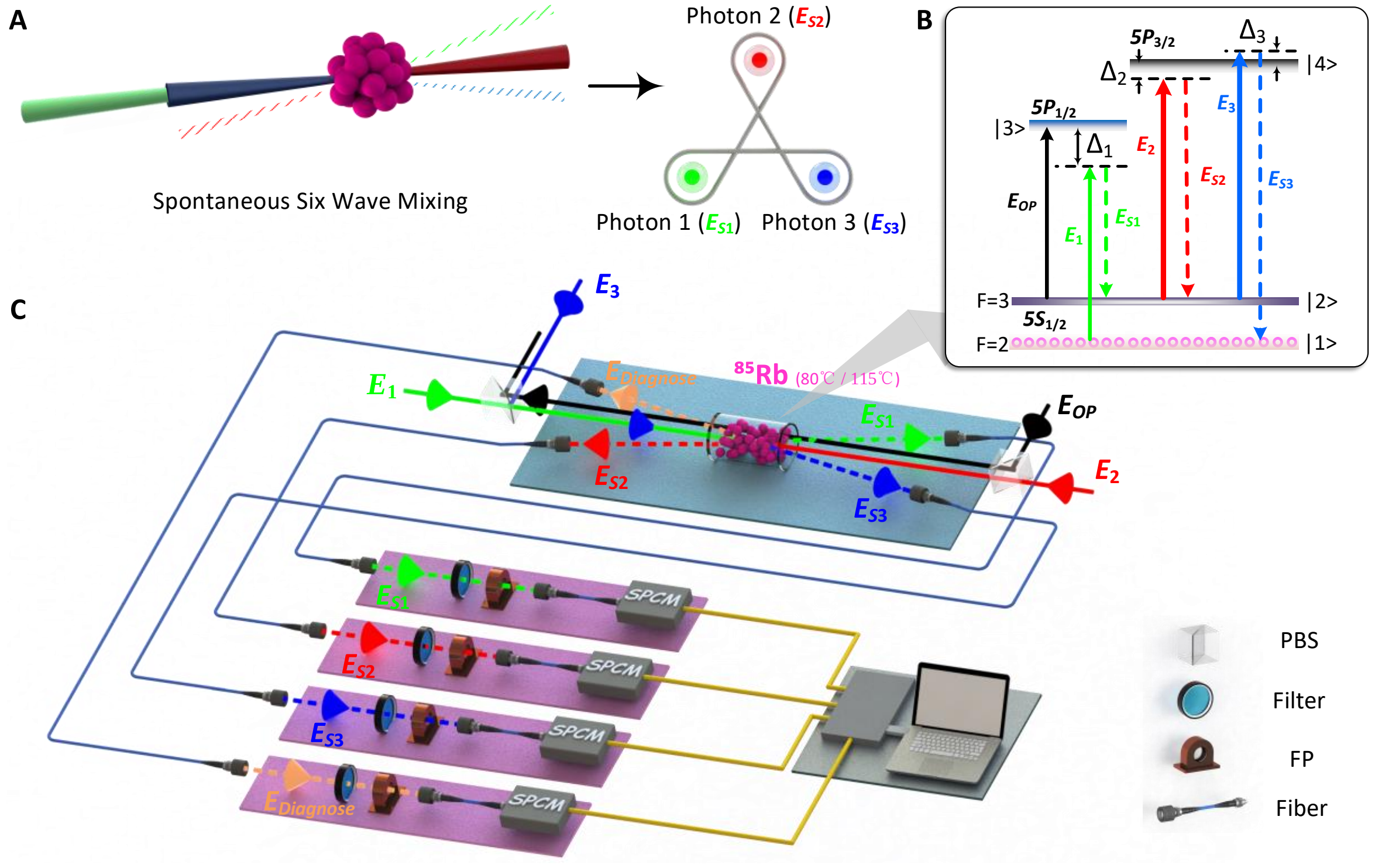

**Fig. 1. Generation of genuine W-triphotons entangled in time-energy directly via SSWM in a hot atomic vapor.** (**A**) Conceptual schematic of creating a W-triphoton state via the fifth-order parametric nonlinear process. (**B**) The [85]Rb energy-level diagram of the SSWM process. (**C**) The experimental setup. Three coaxial input driving fields $E_1$ (795 nm), $E_2$ (780 nm) and $E_3$ (780 nm) are coupled into the center of an [85]Rb vapor cell heated at 80°C (or 115°C) to initiate the simultaneous generation of W-triphotons in $E_{S1}$, $E_{S2}$ and $E_{S3}$. An additional optical-pumping beam $E_{\mathrm{OP}}$ is added to clean up the residual atomic population in the level |2> for preventing the noise from the Raman scattering. The generated photons are coupled into a data acquisition system by single-mode fibers and jointly detected by three synchronized single-photon counting modules (SPCM) with filters (F) and Fabry–Perot cavities (FP) placed in front. To eliminate accidental coincidences caused by dual biphotons and quadraphotons, an extra detection of the diagnosis photons $E_{\mathrm{Diagnose}}$ is applied to ensure the natural triphoton collection. All trigger events are then interrogated by a fast-time acquisition card with a computer.

Physically, the SSWM process can be understood from the effective interaction Hamiltonian

$$H = \epsilon_0 \int_V d^3r \chi^{(5)} E_1 E_2 E_3 E_{S1}^{(-)} E_{S2}^{(-)} E_{S3}^{(-)} + H.c. \ (H.c., \text{Hermitian conjugate}), \quad (1)$$

with three input (output) beams treated as classical (quantized) fields and $V$ being the interaction volume. In Eq. (1), $\chi^{(5)}$ denotes the fifth-order Doppler-broadened nonlinear susceptibility and governs the nonlinear conversion efficiency. In the Schrödinger picture, after some algebra, the triphoton state at the two cell surfaces can be derived from first-order perturbation theory by ignoring the vacuum contribution (SI), and takes the form of

$$|\Psi\rangle \propto \iiint d\omega_{S1} d\omega_{S2} d\omega_{S3} \chi^{(5)} \Phi\left(\frac{\Delta kL}{2}\right) \delta(\Delta\omega) \left|1_{\omega_{S1}}, 1_{\omega_{S2}}, 1_{\omega_{S3}}\right\rangle. \quad (2)$$

Here, $\Delta\omega = \sum_{j=1}^{3}(\omega_{Sj} - \omega_j)$, $L$ is the interaction length, $\Delta k = \Delta\vec{k} \cdot \hat{z}$ is the phase (or wavenumber) mismatch, the phase-mismatch longitudinal function $\Phi(x) = \text{sinc}(x)e^{-ix}$ ascribes the three-photon natural spectral width arising from their different group velocities. Besides conditioning the triphoton output rate, the $\chi^{(5)}$-resonance profile also specifies the generation mechanism along with the photon intrinsic bandwidths. Overall, the state (2) outlines a few peculiar features yet to be experimentally verified: First, because of its non-factorization, $|\Psi\rangle$ is entangled in frequency (or time), instead of polarization. Second, characterized by two independent variables, $|\Psi\rangle$ conforms to the essential characteristics of the tripartite W class, that is, by tracing one photon away, partial entanglement still exists in the remaining bipartite subsystem. Third, since the triphoton waveform is defined by the convolution of $\Phi$ and $\chi^{(5)}$, two distinct types of Glauber third-order (as well as conditional second-order) temporal correlations are expected to be manifested in threefold (and conditioned twofold) coincidence counting measurement. Consequently, two very differing scenarios are expected to be revealed in triphoton coincidence counting measurement. Last, but not the least, the triplet production rate is linear in the intensity of each input laser and can be dramatically enhanced by orders of magnitude by optimizing system parameters. It is worth pointing out that all these striking properties have been well affirmed in our series of experiments. Of importance, this is the first experimental proof of the time-energy-entangled triphoton W state discovered a decade ago (*36*) but never realized.

*Experimental set up*. In experiment, we optimized the SSWM phase-matching condition via controlling the frequency detunings and incident angles of three driving fields so as to effectively collect emitted triphotons. Upon triggering $SPCM_j$, the temporal correlation was concealed in photon counting histograms saved in a fast-time acquisition card with 0.0244-ns bin width, where, within in every time window of 195 ns, the detection of an $E_{S1}$-photon triggered the start of a coincidence event that ended with the detection of subsequent $E_{S2}$- and $E_{S3}$-photons. In most measurements, we collected the total trigger events over an hour and then analyzed the corresponding three-photon coincidences from the histogram in the parameter space $(\tau_{21}, \tau_{31})$, where $\tau_{21} = \tau_2 - \tau_1$ and $\tau_{31} = \tau_3 - \tau_1$ are respectively the relative time delays with $\tau_j$ being the triggering time of the SPCM$_j$.

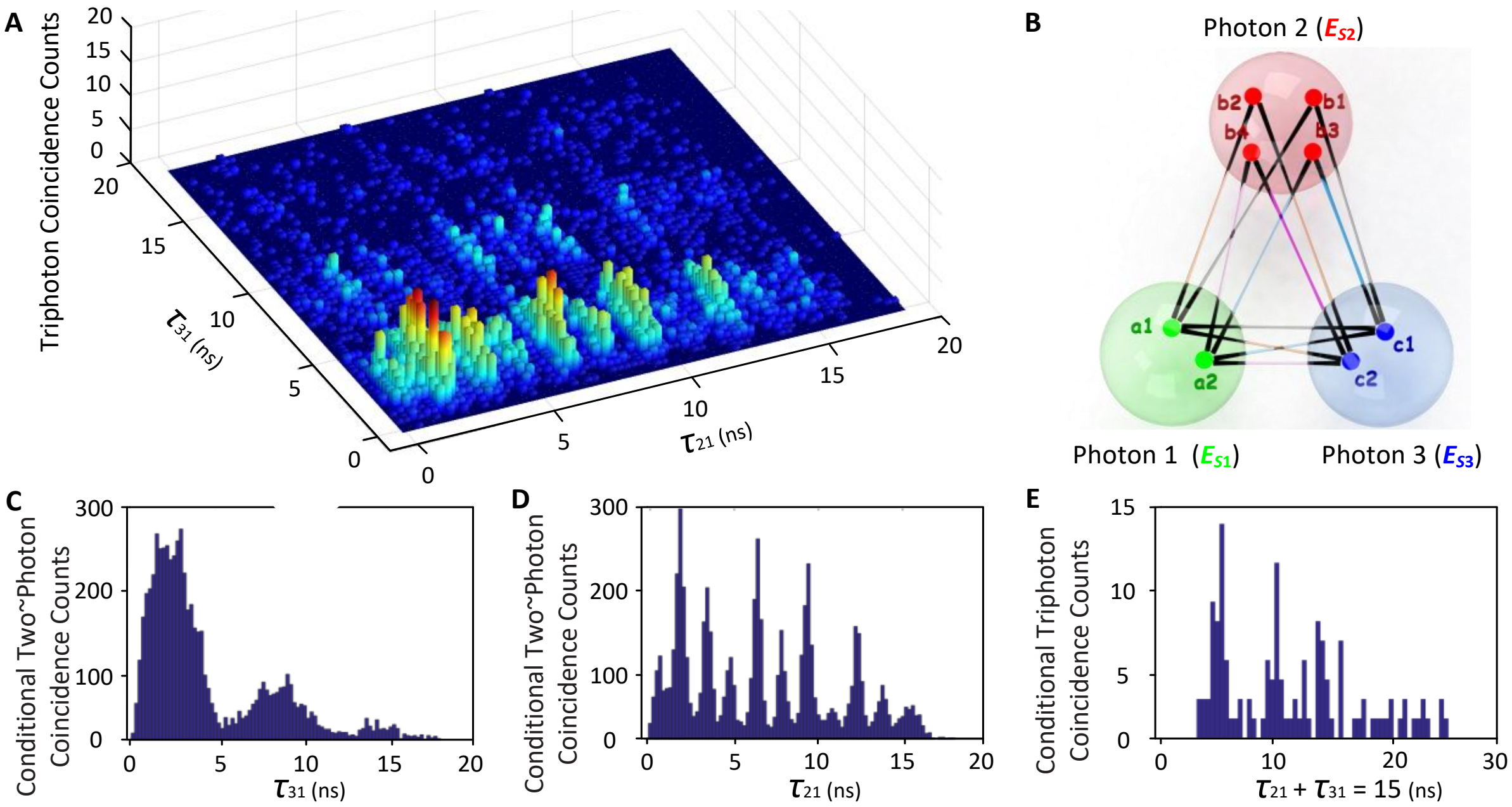

**Fig. 2. Triphoton coincidence counting measurements.** (**A**) Three-dimensional (3D) quantum interference formed by three-photon coincidence counts collected in 1 h with the time-bin width of 0.25 ns for OD = 4.6. The generation rate and accidentals are respectively $102 \pm 9$ and $6 \pm 1$ per minute. The powers of the input $E_1$, $E_2$ and $E_3$ beams are $P_1 = 4$ mW, $P_2 = 40$ mW, and $P_3 = 15$ mW, respectively, and the corresponding frequency detunings are $\Delta_1 = -2$ GHz, $\Delta_2 = -150$ MHz, and $\Delta_3 = 50$ MHz. (**B**) Schematic illustration of triphoton interference originating from the coexistence of multi-SSWMs. (**C**) & (**D**) Conditional two~photon coincidence counts as the function of $\tau_{21}$ and $\tau_{31}$ in (**A**) by tracing the third photon $E_{S3}$ and $E_{S2}$, respectively. (**E**) Conditional three-photon coincidence counts along the trajectory of $\tau_{21} + \tau_{31} = 15$ ns in (**A**).

*Triphoton temporal correlation*. As an exemplar of such, Fig. 2A displays one set of measured threefold coincidence counts from one recorded histogram after subtracting the accidental noise, giving rise to an intriguing three-dimensional temporal correlation with the 18.6- and 19.0-ns effective measurement time window along the $\tau_{21}$- and $\tau_{31}$-axis because of the employed dectors. For the 0.25-ns time-bin width per detector, integrating all involved time bins yields the total of ~$6 \times 10^3$ threefold trigger events, which result in a raw triphoton generation rate of $102 \pm 9$ per minute without account of the coupling loss and detection efficiency. This rate is orders of magnitude higher than any previous one, and can be further improved by applying more efficient SPCMs as well as optimizing the fiber coupling efficiency. From the raw data, the background accidentals were estimated to be $6 \pm 1$ per minute, mainly originating from the residual dual pairs as well as accidental coincidences of uncorrelated singles and dark counts of the SPCMs. This low background noise implies that the undesired third-order nonlinear processes were well filtered out in the experiment. On the other hand, the complicated pattern is a direct consequence of nontrivial W-triphoton interferences due to the occurrence of multiple coexisting SSWM processes in the regime of damped Rabi oscillations. As described previously, these processes arise from the multi-resonance structure of $\chi^{(5)}$. According to our *qualitative* dressed-state calculations (SI), there are four such coexisting channels, as schematic in Fig. 2B, coherently contributing to the observed quantum interference. To confirm that the emitted triphoton state belongs to the W class, we then used the acquired data to investigate the correlation properties of different bipartite subsystems. To do so, we integrated the coincidence counts by tracing away one photon from every triphoton event over that photon's arrival time. In this way, we acquired the conditional two~photon temporal waveforms with $\tau_{21}$ or $\tau_{31}$ as variables, and plotted them, respectively, in Figs. 2C and D. Interestingly, the conditioned $\tau_3$-waveform in Fig. 2D exhibits a damped periodic oscillation with a period of ~6.2 ns (SI); while the $\tau_{21}$-waveform in Fig. 2C reveals two superimposed damped periodic oscillations with another 1.7-ns period in addition to the 6.2-ns one (SI), an interference effect unusual to any existing biphoton source. In contrast, the triphoton waveform has flexible temperal widths, for instance, 28 ns along the direction of $\tau_{21} + \tau_{31} = 15$ ns (Fig. 2E). This contrasting phenomenon also supports our theoretical picture from alternative aspect, that the observed interference is caused by at least three sets of coherently coexisting SSWM processes. As demonstrated in SI, our qualitative analysis gives a good account of the experimental data.

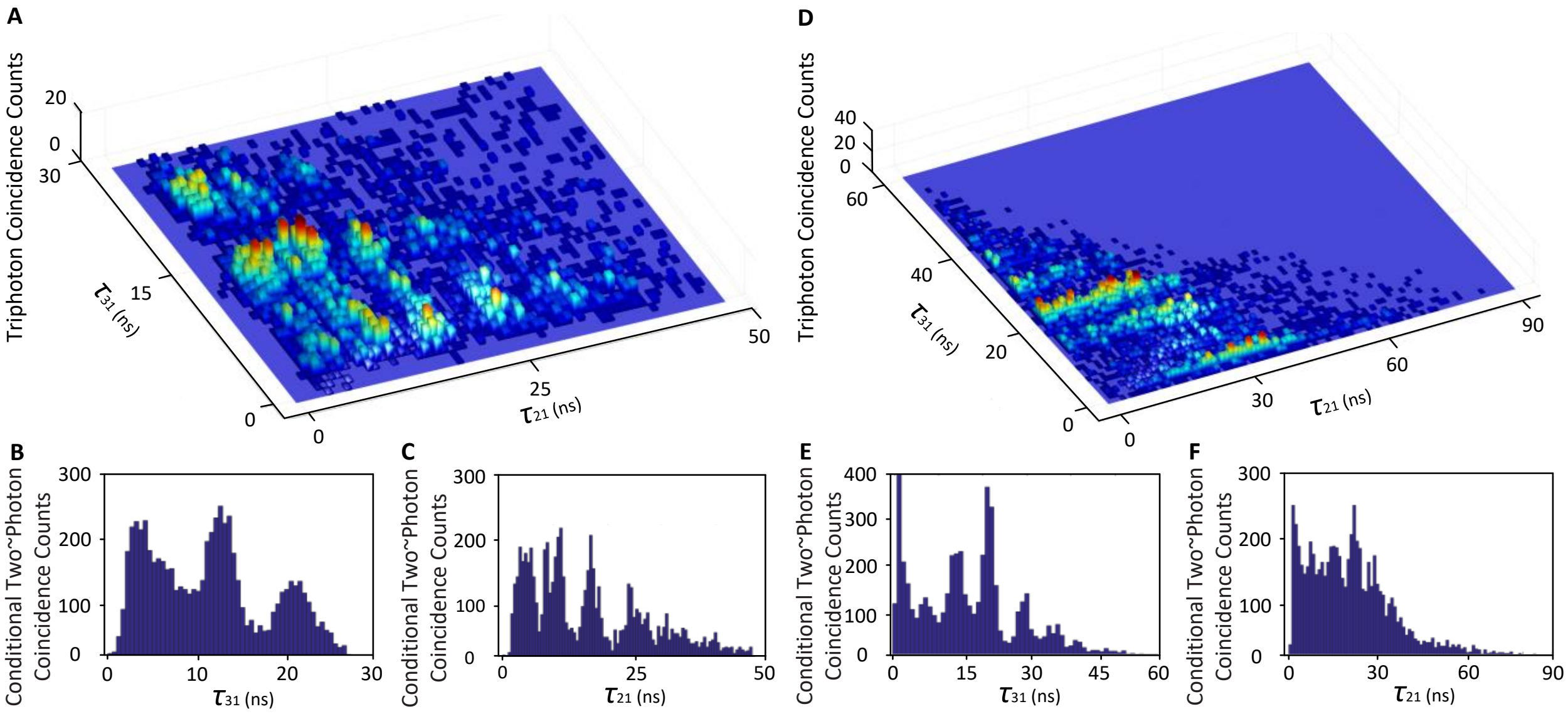

**Fig. 3. Triphoton coincidence counting measurements by tuning the coupling strength and OD.** (**A**) 3D quantum interference formed by three-photon coincidence counts collected in 1h with the 0.7-ns time-bin width by changing $P_2$ to 15 mW and $\Delta_2$ to $-50$ MHz. Other parameters are same as Fig. 2. The generation rate and accidentals rate are 77.4 ± 7.8 and 11 ± 2.1 per minute, respectively. (**B**) & (**C**) Conditional two~photon coincidence counts as the function of $\tau_{21}$ and $\tau_{31}$ in (**A**) by tracing the third photon $E_{S3}$ and $E_{S2}$, respectively. (**D**) Collected over 40 min with 1-ns time-bin width by changing OD to 45.7. Other parameters are same as Fig. 2. The generation and accidentals rates are 125 ± 11 and 28 ± 6.4 per minute, respectively. (**E**) & (**F**) Conditional two~photon coincidence counts as the function of $\tau_{21}$ and $\tau_{31}$ in (**D**).

Since the attributes of triphoton waveforms are dependent on the system parameters, this prompts us to manipulate and control their quantum correlations by means of tuning the input lasers as well as the atomic density or optical depth (OD). To this end, we carried out a series of experiments to tailor temporal correlation by shaping their waveforms by varying various parameters. Two sets of such representative experimental data are presented in Fig. 3. In comparison to Fig. 2A, Fig. 3A shows the steered waveform by reducing the power and frequency detuning of the input $E_2$ laser. As one can see, the profile of the triphoton temporal correlation is dramatically changed in spite of the reduced generation rate $77.4 \pm 7.8$ minute$^{-1}$. Especially, the conditional two~photon coincidence counts manifest mono-periodic oscillations with the same period of 6.2 ns along both $\tau_{21}$ and $\tau_{31}$ directions, as illustrated in Figs. 3B and C. This is because, in this case, the Rabi frequency of $E_2$ was tuned to be very close to that of $E_3$. As a consequence, half of the multiple resonances associated with the emission of $E_{S2}$-photons (Fig. 2B) become degenerate and share the same spectrum. Likewise, the triphoton temporal coherence length along the $\tau_{21} + \tau_{31} = 29$ ns direction is enlarged to 40 ns. On the other hand, triphoton interference can be also modulated by altering the phase-mismatch longitudinal function $\Phi$ in Eq. (2). Akin to the biphoton generation, the phase mismatch $\Delta k$ in $\Phi$ is determined by the linear susceptibility of each mode in SSWM via the EIT slow-light effect. As showcased in Fig. 3D, by augmenting the OD from 4.6 to 45.7, the triphoton temporal correlation is considerably modified by the dispersion relation of the atomic vapor and falls into the group-delay regime. In addition to raising the production rate to $125 \pm 11$ per minute, the oscillatory curvature is markedly suppressed and replaced by the overall decay envelopes. This transformation becomes more evident when examining the conditioned two~photon coincidence counts. By comparing Fig. 3F with Figs. 3B, C and E, one can see that the enhanced dispersion apparently smears the damped Rabi oscillations along the $\tau_{21}$-direction, implying that the narrower bandwidths defined by $\Phi\left(\frac{\Delta k L}{2}\right)$ regulate the bandwidths dictated by $\chi^{(5)}$ to obscure the interference amongst four sets of coexisting SSWM channels. Besides, the triphoton temporal coherence length along the direction of $\tau_{21} + \tau_{31} = 50$ ns is also significantly prolonged up to 70 ns.

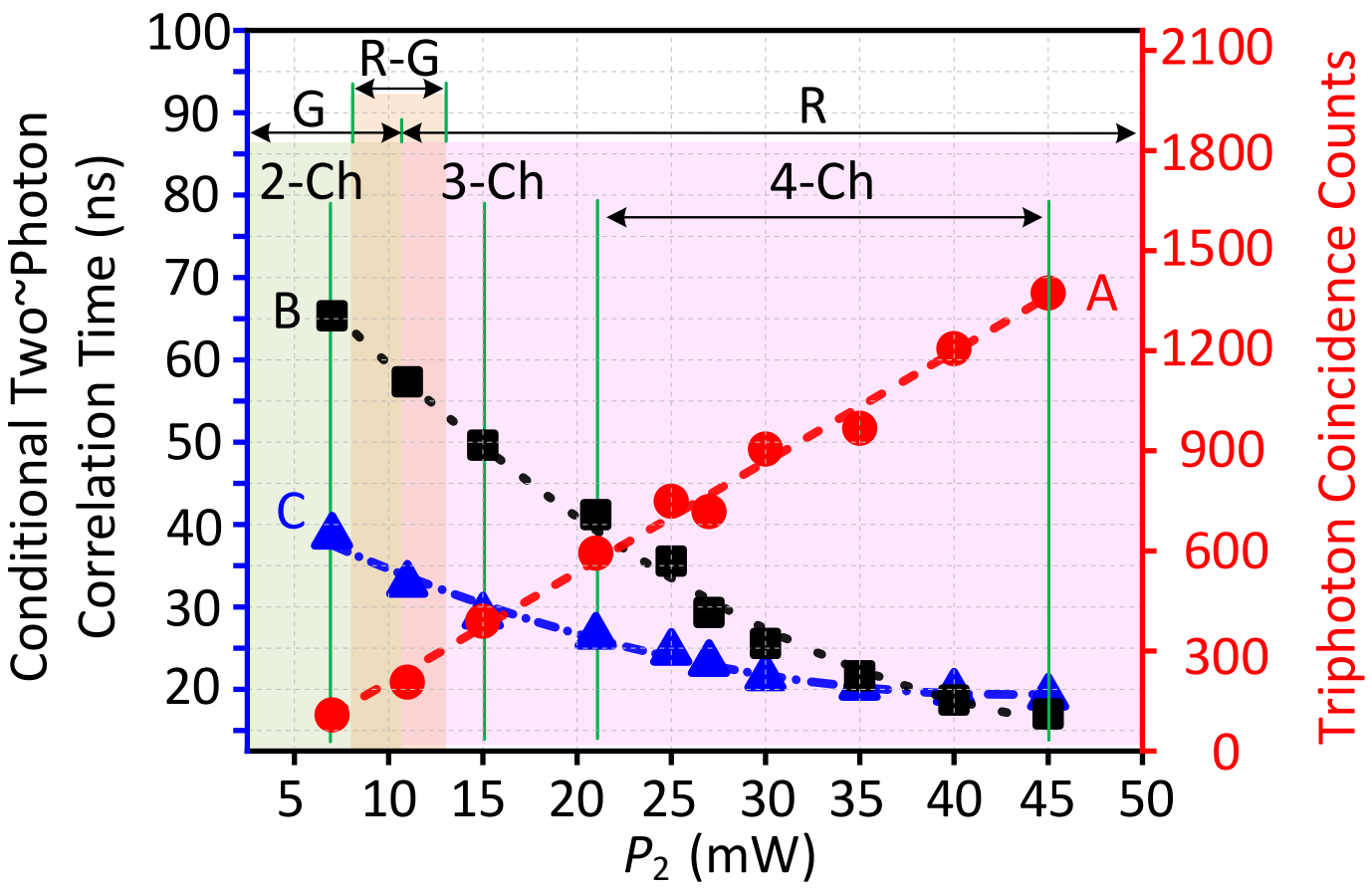

**Fig. 4. Controllable waveform generation.** The triphoton generation rate (red dots) in 15 minutes versus the input power $P_2$ of the driving field $E_2$. The correlation times of conditional two~photon coincidences along the $\tau_{21}$ (black squares) and $\tau_{31}$ (blue triangles) directions by changing $P_2$. By increasing $P_2$, the triphoton temporal correlation is shifted from the group-delay (G) regime to the Rabi-oscillation (R) region. $j$-Ch ($j = 2,3,4$) means the coherent coexistence of $j$ types of indistinguishable SSWMs. The experimental condition is same as that in Fig 2.

To reveal the nonclassicality of the W triphoton state, we continued to examine the violation of the Cauchy-Schwarz inequality (*37*, *38*)as well as the fringe visibilities of the observed Rabi oscillations. By normalizing the threefold coincidence events to the flat background counts along with the additional auto-correlation measurement of the collected $E_{S1}$, $E_{S2}$ and $E_{S3}$ photons, we found that the Cauchy-Schwarz inequality is violated by a factor of $250 \pm 55$ in Fig. 2A, $154 \pm 43$ in Fig. 3A, and $79 \pm 21$ in Fig. 3D. Note that here these values were optimized by filtering possible biphoton processes in measurement. Additionally, we observed that the fringe visibility of Fig. 2A can be as high as $90 \pm 5\%$.

In addition to the above experiments, it is also instructive to explore the triphoton production rate and temporal correlation width as a function of the input pump power for further understanding the proposed generation mechanism. This has motivated us to implement additional measurements and the experimental data is presented in Fig. 4. As one can see, indeed, the triphoton generation rate follows a linear growth in the input power $P_2$ of the $E_2$ field. For the temporal coherence length, we concentrated on the two~photon conditional coincidence counting along the $\tau_{21}$ and $\tau_{31}$ directions. From Fig. 4, it is not difficult to find that increasing $P_2$ results in the reduction of the correlation time. This stems from the reduced slow-light effect when augmenting $P_2$. Note that Figs. 2A, 3A and 3D simply become one individual point in Fig. 4. Overall, our approach enables all-optical coherent manipulation to create the genuine triphotons with controllable waveforms.

## Discussion

In conclusion, we have for the first time observed the efficient dependable continuous-mode W-triphoton emission directly through SSWM in a warm atomic vapor with a generation rate of about $125 \pm 11$ min$^{-1}$. Moreover, due to the coexistence of multi-SSWMs, these time-energy-entangled W triphotons have resulted in various nontrivial three-photon temporal interferences. Furthermore, by manipulating the system parameters, the triphoton temporal correlations can be flexibly engineered and tailored and demonstrate many peculiar characteristics inaccessible to all previous mechanisms. As a reliable source, it is expected to play a vital role in probing foundations of quantum theory and advancing various quantum-based technologies in information processing, communications, networking, imaging, metrology, etc.

## Methods

**Experimental implementation.** Experimentally, three coaxial driving beams $E_1$, $E_2$ and $E_3$ are coupled to the center of the $^{85}$Rb vapor cell to initiate the SSWM process, as shown in Fig. 2. The relevant energy-level diagram is shown in Fig. 1B, where the atoms are prepared at the ground level $|1\rangle$ ($5S_{1/2}, F = 2$). The other involved energy levels are $|2\rangle$ ($5S_{1/2}, F = 3$), $|3\rangle$ ($5P_{1/2}$), and $|4\rangle$ ($5P_{3/2}$). The horizontally polarized weak probe $E_1$ beam at the 795-nm wavelength is applied the atomic transition $|1\rangle \to |3\rangle$ with a large red frequency detuning $\Delta_1$ (2 GHz) so that the atomic population resides primarily at $|1\rangle$. The other two strong coupling beams $E_2$ (780 nm, horizontal polarization) and $E_3$ (780 nm, vertical polarization) are near resonantly coupled to the same atomic transition $|2\rangle \to |4\rangle$ but with changeable detunings $\Delta_2$ and $\Delta_3$. By carefully adjusting the phase matching conditions, the spatially separated triphotons $E_{S1}$, $E_{S2}$ and $E_{S3}$ with wave vectors $\vec{k}_{S1}$, $\vec{k}_{S2}$ and $\vec{k}_{S3}$ are spontaneously emitted along the phase-matching directions with a small forward angle about 4° away from the three driving fields. Besides, we have added an additional optical-pumping beam $E_{OP}$ to clean up the residue atomic population in $|2\rangle$ so that the Raman scattering can be suppressed from the transition $|2\rangle \to |3\rangle$. To increase the fifth-order nonlinearity, the $^{85}$Rb vapor cell with a length of $L = 7$ cm is heated to 80°C (or 115°C). In this regard, the reported data in Figs. 2 and 3A-C were collected at the temperature of 80°C; while the data presented in Figs. 3D-F were obtained at 115°C. Also, the narrowband filters and customized interference etalon Fabry-Perot (FP) cavities are placed in front of each SPCM to filter the scattered driving lasers from the collected triphoton trigger events. After detected by SPCMs, the trigger events are recorded by a time-to-digit converter, where the maximum resolution time of our recording card is 813 fs. In our experiment, the fiber-fiber coupling efficiency and the SPCM detection efficiency are 70% and 40%, respectively.

**Filtering possible biphoton processes from triphoton coincidence counts.** Although the triphoton generation by SSWM is the focus of the measurement, due to the larger magnitude of the third-order nonlinearity, it is necessary to consider the possible false counts from the biphoton processes. Based on the atomic level structure and the adopted field coupling geometry, there are seven crucial SFWMs (Fig. S6 in SI) that may result in accidental coincidences: (1) SFWM1 initiated by $E_1$ and $E_2$, (2) SFWM2 by $E_1$ and $E_3$, (3) SFWM3 by $E_2$ and $E_3$, (4) SFWM4 by $E_3$ and $E_2$, (5) SFWM5 by $2E_1$, (6) SFWM6 by $2E_2$, and (7) SFWM7 by $2E_3$. Specifically, the biphotons produced from the following SFWMs may contribute to the accidental joint-detection probability: (1) SFWM1 + SFWM2, (2) SFWM1 + SFWM3, (3) SFWM1 +SFWM4, (4) SFWM1 + SFWM5, (5) SFWM1 + SFWM7, (6) SFWM2 + SFWM3, (7) SFWM2 + SFWM4, (8) SFWM2 + SFWM6, (9) SFWM3 + SFWM4, (10), SFWM3 + SFWM5, (11) SFWM3 + SFWM7, (12) SFWM4 + SFWM5, (13) SFWM4 + SFWM7, (14) SFWM5 + SFWM6, and (15) SFWM6 +SFWM7. Fortunately, the central frequency difference of the similar photons from SSWM and SFWMs are more than 3 GHz. Therefore, before being detected by SPCMs, the collected photons need to pass through the high-quality single-frequency band filters and the customized narrowband etalon Fabry-Perot cavity (with a bandwidth ~600 MHz). The bandwidth, transmission efficiency, and extinction ratio of the employed filters are 650 MHz, 80%, and 60 dB, respectively. After these measures, most of the biphoton noise can be filtered from the detection. In addition, the phase-matching condition for the SSWM process is much different from those for the possible SFWM processes. For instance, the photons from SFWM2 have distinctive emission angles from those from SSWM. As a result, the three-photon coincidence counts in actual measurements are mainly determined by true triphotons, uncorrelated singles, and dark counts. In practice, the biphotons and uncorrelated singles can be well filtered in the three-photon coincidence counting measurement by carefully adjusting the phase-matching conditions.

**Additional detection of diagnose photons $\boldsymbol{E}_{\mathbf{Diagnoise}}$.** To further guarantee the detected photons that are really from SSWM, we have performed one additional detection of the two-photon coincidences $E_{S3}$ and $E_{\text{Diagnose}}$ simultaneously in conjunction with the coincidences between $E_{S1}$ and $E_{S2}$ by artificially introducing the diagnose photons $E_{\text{Diagnose}}$. This arrangement allows us to greatly reduce the false three-photon trigger events from dual biphotons particularly. The experimental results of $E_{S3}$ and $E_{\text{Diagnose}}$ are given in the SI. By the same reconstruction method, we notice that the trigger events from two pairs of biphotons can be safely removed from the data recording.

**The Cauchy-Schwarz inequality.** The nonclassicality of triphoton correlation can be verified by observing the violation of the well-known Cauchy-Schwarz inequality, which is defined by

$$\frac{\left[g^{(3)}(\tau_{21},\tau_{31})\right]^2}{\left[g_{S1}^{(1)}\right]^2\left[g_{S2}^{(1)}\right]^2\left[g_{S3}^{(1)}\right]^2} \leq 1.$$

Here, $g^{(3)}(\tau_2,\tau_3)$ is the normalized third-order correlation function with respect to the accidental background. $g_{S1}^{(1)}$, $g_{S2}^{(1)}$ and $g_{S3}^{(1)}$ are the normalized autocorrelations of the emitted photons $E_{S1}$, $E_{S2}$ and $E_{S3}$ measured by a fiber beam splitter. In our experiment, the nonzero background floor in such as Figs. 2 and 3 is a result of the accidental coincidences between uncorrelated single photons. According to the measured data, we estimate that the maximum values of $g_{S1}^{(1)}$, $g_{S2}^{(1)}$ and $g_{S3}^{(1)}$ are respectively to be $1.6 \pm 0.2$, 2 and 2.

**Acknowledgments**

We are grateful to Xinghua Li, Dan Zhang, and Da Zhang for their contributions at the early stage of the project and to Yanhua Zhai for helpful discussions on designing the detection system.

**Funding:**

National Key Research and Development Program of China (2017YFA0303700, 2018YFA0307500),

Key Scientific and Technological Innovation Team of Shaanxi Province (2021TD-56),

National Natural Science Foundation of China (61975159, 12174302, 62022066, 12074306, 12074303).

Wen was supported by NSF 2329027.

# Supplementary Materials for

## Direct generation of time-energy-entangled W triphotons in atomic vapor

Kangkang Li,[1] Jianming Wen,[2]* Yin Cai,[1]* Saeid Vashahri Ghamsari[2], Changbiao Li[1], Feng Li[1], Zhaoyang Zhang[1], Yanpeng Zhang[1*], and Min Xiao[3,4]

*Corresponding author. Email: jianming.wen@kennesaw.edu; caiyin@xjtu.edu.cn; ypzhang@mail.xjtu.edu.cn.

**This PDF file includes:**

Supplementary Text
Figs. S1 to S13
References (*9*, *27*, *31*-*33*, *39*-*57*)

**Supplementary Text**

I. Qualitative Theory of Time-Energy-Entangled W Triphoton Generation in Atomic Vapor

Qualitative Derivation of Fifth-Order Nonlinear Susceptibility $\chi^{(5)}$

Nonlinear optics stands as a foundational pillar in the realm of generating, shaping, and transforming quantum light. In the pursuit of harnessing nonclassical light through the deployment of atomic ensembles, the optical response of these systems, encompassing both linear and nonlinear susceptibilities, emerges as a pivotal determinant shaping the characteristics of the resultant quantum states and waveforms. This influence is particularly pronounced when the interaction between light and atoms transpires in proximity to atomic resonance, and the nonclassical light generated is notably weaker than the input driving fields. Consequently, a fundamental challenge inherent in such scenarios is the derivation of linear and nonlinear susceptibilities governing the interplay of the involved electromagnetic (EM) fields.

In the realm of optical interactions, the computational landscape for determining susceptibilities becomes more intricate when dealing with scenarios involving multiple EM fields acting on the same atomic transition. In cases where only one EM field per atomic transition is implicated, established methods such as density-matrix formalism and master equations prove effective in calculating susceptibilities. Yet, as the complexity deepens, especially in the context of triphoton generation as examined in this study, novel strategies are necessitated.

Wen and colleagues have contributed a valuable approach (*31-33*) that facilitates precise susceptibility calculations, particularly relevant for generating entangled photon pairs. However, when applied to the triphoton generation investigated herein, the methodology faces heightened theoretical calculations. This complexity arises from the simultaneous presence of three EM fields–$E_2$, $E_3$, and $E_{s2}$–within a single atomic transition $|2\rangle - |4\rangle$ (as depicted in Fig. 1B in the main text). Ongoing efforts are dedicated to advancing the exact derivations using this method.

In the interim, we employ a "qualitative" technique–perturbation chain rule–to explore the optical response of atomic vapor in the context of triphoton emission and its associated optical attributes. This qualitative approach has found application in analogous atomic systems with comparable energy-level structures, yielding results that align comparably. Moreover, it has been employed to analyze light-atom interaction (*39-44*) in the context of six-wave mixing (SWM) in the stimulated emission regime. As elucidated below, while *the derived qualitative optical response results* may not align seamlessly with the experimental data, they do furnish a reasonable framework for comprehending the observed triphoton behaviors.

The foundation of this qualitative approach is firmly grounded in perturbation theory, which prioritizes the dressing steady states while overlooking the transient propagation influence. The initial stage involves perturbative examination of the SWM process, leveraging the framework of weak-field approximation. Subsequently, the dressing perturbation strategy is invoked, establishing a set of strongly coupled equations driven by the strong fields. This framework thus facilitates the determination of density-matrix elements via the perturbation chain rule.

Following the methodology akin to that expounded in Refs. (*39-44*), it is revealed that the fifth-order nonlinear susceptibility $\chi^{(5)}$ can be approximately attained from the ensuing perturbation chain:

$$\rho_{11}^{(0)} \xrightarrow{\omega_1} \rho_{31}^{(1)} \xrightarrow{\omega_{S1}} \rho_{21}^{(2)} \xrightarrow{\omega_2} \rho_{41}^{(3)} \xrightarrow{\omega_{S2}} \rho_{11}^{(4)} \xrightarrow{\omega_3} \rho_{41}^{(5)}, \quad \text{(S1)}$$

where $\omega_1$, $\omega_2$ and $\omega_3$ denote the frequencies of the three input lasers, while $\omega_{S1}$, $\omega_{S2}$ and $\omega_{S3}$ represent the frequencies of the generated triphotons. By solving the series of density-matrix equations, one can deduce the density-matrix elements $\rho_{11}^{(0)}$, $\rho_{31}^{(1)}$, …, $\rho_{41}^{(5)}$ in Eq. (S1) through a stepwise progression. Given the nature of atomic vapor, the incorporation of Doppler broadening effects become imperative. After some lengthy calculations, we have finally derived the fifth-order nonlinear susceptibility characterizing the light-atom interaction, as displayed in Fig. S1. This susceptibility adopts the following form:

$$\chi^{(5)}(\delta_2,\delta_3) = \int_{-\infty}^{\infty} dv \frac{2N\mu_{13}\mu_{24}\mu_{23}\mu_{14}^3 f(v)}{\varepsilon_0\hbar^5\left\{\begin{array}{c}(\Gamma_{31}+i\Delta_{D1})[(\Gamma_{21}+iW_{D-}\delta_2+iW_{D+}\delta_3)(\Gamma_{41}+iW_{D-}\delta_2+iW_{D+}\delta_3+i\Delta_{D2})+|\Omega_2|^2] \\ \times[(\Gamma_{11}+iW_{D+}\delta_3)(\Gamma_{41}+iW_{D+}\delta_3+i\Delta_{D3})+|\Omega_3|^2]\end{array}\right\}}. \quad \text{(S2)}$$

Here, $f(v) = \sqrt{\frac{m_{\mathrm{Rb}}}{2\pi k_B T}} e^{-\frac{m_{\mathrm{Rb}} v^2}{2k_B T}}$ is the Maxwell-Boltzmann velocity distribution of Rb atoms in thermal motion, where $m_{\mathrm{Rb}}$ is the mass of the Rb atom, $k_B$ is the Boltzmann constant, $T$ is the vapor temperature, and $v$ is the atomic kinetic velocity; $N$ denotes the atomic density; $\mu_{ij}$ $(i,j = 1,2,3,4)$ represents the electric dipole matrix element for the atomic transition $|i\rangle \to |j\rangle$; $\varepsilon_0$ stands for the vacuum permittivity; $\Gamma_{ij}$ is the decay or decoherence rate between levels $|i\rangle$ and $|j\rangle$; $\Delta_{D1} = \Delta_1 + v\omega_{31}/c$, $\Delta_{D2} = \Delta_2 - v\omega_{42}/c$, and $\Delta_{D3} = \Delta_3 + v\omega_{42}/c$ are associated with the frequency detunings $\Delta_1 = \omega_{31} - \omega_1$, $\Delta_2 = \omega_{42} - \omega_2$, and $\Delta_3 = \omega_{42} - \omega_3$, where $\omega_{ij}$ is the frequency difference between $|i\rangle$ and $|j\rangle$; $W_{D\pm} = 1 \pm v/c$ depends on atomic velocity with $c$ the speed of light in vacuum; $\Omega_2$ and $\Omega_3$ are the Rabi frequencies; $\delta_2$ and $\delta_3$ define the spectral distributions with respect to the central frequencies of the emitted $E_{S2}$ and $E_{S3}$ photons, respectively. Additionally, it's important to note that owing to the energy conversation in SSWM, the triggers for these two photons require the detection of the output $E_{S1}$ photon at frequency $\omega_{S1} = \omega_1 + \omega_2 + \omega_3 - \omega_{S2} - \omega_{S3}$. This alternatively implies that the spectral distributions of the entangled three-photon state need to satisfy the condition $\delta_1 + \delta_2 + \delta_3 = 0$.

When $T = 80°\mathrm{C}$, the Doppler width is estimated to be approximately $\Delta_D$= 555 MHz, orders of magnitude larger than the Rb natural linewidth. The atomic density is given by $N = 1.2 \times 10^{11}$ cm$^{-3}$. The optical depth (OD), calculated as $OD = N\sigma_{41}L$, amounts to 4.6, where $\sigma_{41} = \frac{\omega_{41}|\mu_{14}|^2}{2\varepsilon_0\hbar c\Gamma_{41}\Delta_D} = 3\pi N\Gamma_{41}c^2L/2\Delta_D\omega_{41}^2$ stands for the on-resonance absorption cross-section of the transition $|1\rangle \to |4\rangle$. At a higher temperature $T = 115°\mathrm{C}$, the OD grows significantly to the value of 45.7.

In accordance with our previous theoretical investigations (*9*, *27*, *31-33*, *39-50*), the temporal correlations inherent in the triphoton generation are impacted by two primary factors: the spectral profile of the fifth-order nonlinear susceptibility $\chi^{(5)}$, as provided by Eq. (S2), and the longitudinal phase-mismatch function, which will be discussed in subsequent sessions. With this premise, we initiate our examination by delving into the at the structure of $\chi^{(5)}$.

Similar to our earlier analyses (*9*, *27*, *31-33*, *39-49*), the resonances originating from the denominator of $\chi^{(5)}$ in Eq. (S2) are centrally located around $\delta_{1\pm} = (\Delta_{D2} \pm \Omega_{E_2})/2(1 - \frac{v}{c})$, $\delta_{2\pm\pm} = (\Delta_{D3} - \Delta_{D2} \pm \Omega_{E_2} \pm \Omega_{E_3})/2(1 + \frac{v}{c})$, and $\delta_{3\pm} = (-\Delta_{D3} \pm \Omega_{E_3})/2(1 - \frac{v}{c})$. Here, the effective Rabi frequencies are redefined as $\Omega_{E_2} = \sqrt{\Delta_{D2}{}^2 + 4|\Omega_2|^2 + 4\Gamma_{21}\Gamma_{41}}$ and $\Omega_{E_3} = \sqrt{\Delta_{D3}{}^2 + 4|\Omega_3|^2 + 4\Gamma_{11}\Gamma_{41}}$, with $\Omega_2$ and $\Omega_3$ being the original Rabi frequencies of the $E_2$ and $E_3$ fields, respectively. Notably, the effective linewidths of these resonances are determined by the imaginary components of the denominator. These linewidths, $\Gamma_{\delta_2} = \frac{\Gamma_{21}+\Gamma_{41}}{2} + \frac{\Gamma_{21}\Delta_{D2}}{\Delta_{D2}+\Omega_{E_2}}$ and $\Gamma_{\delta_3} = \frac{\Gamma_{11}+\Gamma_{41}}{2} + \frac{\Gamma_{11}\Delta_{D3}}{\Delta_{D3}+\Omega_{E_3}}$, are responsible for setting the temporal correlation lengths between generated triphotons. Importantly, these resonance centers and effective linewidths are both contingent on the velocity of the atomic motion, and are thus influenced by the Doppler broadening effect.

By analyzing the calculated $\delta_{1\pm}$, $\delta_{2\pm\pm}$ and $\delta_{3\pm}$, it is anticipated that there will generally exist four sets of indistinguishable SSWM processes, facilitating the production of time-energy-entangled W-triphotons. As an illustrative instance, Fig. S2 visually presents the behavior of $\chi^{(5)}$ across different scenarios. A keen observation reveals that, upon velocity integration, for cases with low OD, four distinct resonances will typically manifest (Figs. S2A and S2B); whereas for high OD values, the possibility arises to coalesce four resonances into two (Fig. S2C).

<u>Qualitative Derivations of Linear Susceptibilities $\chi$</u>

Apart from the resonance linewidths governed by $\chi^{(5)}$, the temporal correlation of triphotons is also dependent on dispersion, which stems from the linear optical response. By applying the appropriate perturbation chain rule, after some calculations we obtain the individual linear susceptibilities of the new $E_{S1}$, $E_{S2}$ and $E_{S3}$ fields, yielding the following expressions:

$$\chi_{S1} \approx 0, \tag{S3}$$

$$\chi_{S2} = \int f(v) \frac{-i4N\mu_{24}^2\left((1-\frac{v}{c})\delta_2+i\Gamma_{22}\right)}{\varepsilon_0\hbar\left[4\left((1-\frac{v}{c})\delta_2-\Delta_{D2}+i\Gamma_{42}\right)\left((1-\frac{v}{c})\delta_2+i\Gamma_{22}\right)+|\Omega_2|^2\right]} dv, \tag{S4}$$

$$\chi_{S3} = \int f(v) \frac{-i4N\mu_{14}^2\left((1+\frac{v}{c})\delta_3+i\Gamma_{11}\right)}{\varepsilon_0\hbar\left[4\left((1+\frac{v}{c})\delta_3-\Delta_{D3}+i\Gamma_{41}\right)\left((1+\frac{v}{c})\delta_3+i\Gamma_{11}\right)+|\Omega_3|^2\right]} dv. \tag{S5}$$

Eq. (S3) is amply substantiated by the utilization of a weak input $E_1$ beam coupled with an exceedingly large red detuning $\Delta_1 = -2$ GHz from the transition $|1\rangle \rightarrow |2\rangle$. This outcome indicates that the group velocity of the $E_{S1}$ photons closely approximates the speed of light in vacuum, $c$. To enhance understanding, Fig. S3 encompasses numerical simulations of $\chi_{S2}$ and $\chi_{S3}$, elucidating the features of their profiles. Consequently, the group velocities experienced by the $E_{S2}$ and $E_{S3}$ photons are routinely derived using the formula:

$$v_{S2} = \left(\frac{dk_{S2}}{d\omega}\right)^{-1} = \frac{c}{1+\delta_2(\frac{dn_{S2}}{d\delta_2})}, \tag{S6}$$

$$v_{S3} = \left(\frac{dk_{S3}}{d\omega}\right)^{-1} = \frac{c}{1+\delta_3(\frac{dn_{S3}}{d\delta_3})}, \tag{S7}$$

where $n_{S2} = \sqrt{1 + \mathrm{Re}[\chi_{S2}]}$ and $n_{S3} = \sqrt{1 + \mathrm{Re}[\chi_{S3}]}$ are refractive indices. The imaginary parts of $\chi_{S2}$ and $\chi_{S3}$ ascribe the linear Raman gain or absorption undergone by the $E_{S2}$ and $E_{S3}$ photons during their traversal through the medium. Armed with this insight, the computation of the longitudinal phase mismatch in the SSWM process becomes apparent. This mismatch is defined as

$$\Delta k(\delta_2, \delta_3) = k_{S1} - k_{S2} + k_{S3} - k_1 + k_2 - k_3, \tag{S8}$$

where $k_j = \overline{k_j} + \frac{\omega}{v_j}$ $(j = 1,2,3,S1,S2,S3)$, and $\overline{k_j}$ denotes the central wavenumber. Equation (S8) underscores the inherent spectral width of the generated triphoton state, thereby serving as a natural determinant for the temporal coherence time due to the influence of light propagation within the atomic vapor.

To offer insights into the behavior of $\chi_{S2}$ and $\chi_{S3}$, we present an illustrative example in Fig. S3, showcasing their real and imaginary components post the Doppler integration. As one can see, $\chi_{S2}$ and $\chi_{S3}$ typically exhibit two resonance structures, as visualized in Figs. S3A–D. This divergence from the four resonances observed in $\chi^{(5)}$ (depicted in Figs. S2A and S2B) can be attributed to the qualitative model employed for the calculation of linear (and nonlinear) susceptibilities. We are presently engaged in refining this understanding by undertaking precise theoretical computations of both linear and nonlinear optical responses, leveraging the accurate model (*31-33*) pioneered by Wen *et al*. The outcomes of this ongoing effort are slated for publication in an upcoming venue. Meantime, we are open to the emergence of alternative theories from the community, as the associated mathematics is highly complex. We enthusiastically welcome the development of new theories that can accurately characterize these optical responses. We are optimistic that this complexity presents an opportunity for our work to inspire novel theoretical advancements. Unlike previous protocols that comfortably fit within the existing theoretical framework, our approach challenges it and encourages fresh theoretical development.

Derivation of the Triphoton State $|\Psi\rangle$

To calculate the resultant three-photon state stemming from the SSWM process at the output surface of the medium, we shall work in the Schrödinger picture. We commence with the following effective interaction Hamiltonian,

$$H = \int_0^L dz\varepsilon_0\,\chi^{(5)}E_1^{(+)}E_2^{(+)}E_3^{(+)}E_{S1}^{(-)}E_{S2}^{(-)}E_{S3}^{(-)} + H.c., \tag{S9}$$

where $H.c.$ means the Hermitian conjugate. Here, the generated $E_{S1}$, $E_{S2}$ and $E_{S3}$ photons are described by the quantized electric fields,

$$E_{Sj}^{(+)} = \sum_{k_{Sj}} E_{Sj}a_j e^{i(k_{Sj}z-\omega_{Sj}t)}, \tag{S10}$$

where $a_j$ symbolizes the annihilation operator for the mode with the wavenumber $k_{Sj}$ and angular frequency $\omega_{Sj}$. Additionally, $E_{Sj} = i\sqrt{\hbar\omega_{Sj}/2\varepsilon_0 n_{Sj}^2 L}$. On the other hand, the three input continuous-wave (cw) lasers $E_1$, $E_2$, and $E_3$ are taken as classical plane waves,

$$E_1^{(+)} = E_1 e^{i(k_1 z-\omega_1 t)},\ E_2^{(+)} = E_2 e^{i(-k_2 z-\omega_2 t)},\ \text{and}\ E_3^{(+)} = E_3 e^{i(k_3 z-\omega_3 t)}. \tag{S11}$$

The state vector of the triphotons can then be ascertained through first-order perturbation theory (*9, 27, 31-33, 39-51*):

$$|\Psi\rangle = \frac{-i}{\hbar}\int_{-\infty}^{+\infty} dtH\,|0\rangle, \tag{S12}$$

with $|0\rangle$ being the initial vacuum state. By applying Eqs. (S9)–(S12) and ignoring the vacuum term that has no effect in photon clicks, the triphoton state (S12) can be formulated as:

$$|\Psi\rangle = \sum_{k_{S1}}\sum_{k_{S2}}\sum_{k_{S3}} F(k_{S1},k_{S2},k_{S3})\, a_{k_{S1}}^\dagger a_{k_{S2}}^\dagger a_{k_{S3}}^\dagger|0\rangle, \tag{S13}$$

where the three-photon spectral function $F$ is defined as

$$F(k_{S1},k_{S2},k_{S3}) = A\chi^{(5)}\Phi(\Delta kL)\delta(\omega_1+\omega_2+\omega_3-\omega_{S1}-\omega_{S2}-\omega_{S3}), \tag{S14}$$

with $A$ being a grouped constant. In Eq. (S14), the Dirac $\delta$ function comes from the time integral in the steady-state approximation, ensuring the energy conservation in the SSWM process. From the perspective of atomic population, this energy conservation implies that after a triphoton generation cycle, the population returns to its initial ground state $|1\rangle$. $\Phi(\Delta kL)$ is the so-called longitudinal phase-mismatch function, taking the form of:

$$\Phi(\Delta kL) = \frac{1-e^{-i\Delta kL}}{i\Delta kL} = \mathrm{sinc}\left(\frac{\Delta kL}{2}\right) e^{-i\Delta kL/2}. \tag{S15}$$

Due to the Doppler effect in $\chi^{(5)}$ and $\Delta k$, an exact analytical expression for the triphoton state (S13) becomes challenging. Instead, hereafter we will rely on numerical analysis to unveil the triphoton properties.

Derivations of Temporal Correlations of W Triphotons

The optical properties of the W-type triphotons can be comprehensively understood by examining their photon statistics through photon-counting measurements. Consequently, we delve into the temporal correlation of triphotons by evaluating the Glauber second-order and third-order correlation functions. This exploration then prompts us to carry out the analysis of conditioned two~photon coincidence counts and three-photon coincidence counts.

In line with the experimental setup illustrated in Fig. 1 of the main text, the average triphoton coincidence counting rate is expressed as:

$$R_3 = \lim_{T\to\infty} \frac{1}{T}\int_0^T dt_1 \int_0^T dt_2 \int_0^T dt_3 \langle\Psi|\, E_{S1}^{(-)}(\tau_1)E_{S2}^{(-)}(\tau_2)E_{S3}^{(-)}(\tau_3)E_{S3}^{(+)}(\tau_3)E_{S2}^{(+)}(\tau_2)E_{S1}^{(+)}(\tau_1)|\Psi\rangle, \tag{S16}$$

and the conditional two~photon coincidence counting rate is:

$$R_2 = \lim_{T\to\infty} \frac{1}{T}\int_0^T dt_1 \int_0^T dt_2\, \langle\Psi|E_{S2}^{(-)}(\tau_2)E_{S3}^{(-)}(\tau_3)E_{S3}^{(+)}(\tau_3)E_{S2}^{(+)}(\tau_2)|\Psi\rangle, \tag{S17}$$

assuming, for instance, that the $E_{s1}$ photons are traced away. In Eqs. (S16) and (S17), $E_{Sj}^{(+)}(\tau_j)$ ($j = 1,2,3$) is the positive frequency part of the free-space electric field evaluated at the spatial coordinate $r_j$ of the $j$th detector alongside with its trigger (or click) time $t_j$, with $\tau_j = t_j - r_j/c$. For simplicity, we consider the efficiencies of all involved single-photon detectors to be unity. In addition, given that the narrow bandwidths (less than GHz) of the triphotons in question are comparable to or smaller than the spectral resolving width of the utilized single-photon detectors in our experiment, we can simplify Eqs. (S16) and (17) to:

$$R_3 = \left|\langle 0|E_{S3}^{(+)}(\tau_3)E_{S2}^{(+)}(\tau_2)E_{S1}^{(+)}(\tau_1)|\Psi\rangle\right|^2 = |A_3(\tau_1,\tau_2,\tau_3)|^2, \tag{S18}$$

$$R_2 = \sum_{k_{S1}}\left|\langle 0|a_{k_{S1}}E_{S3}^{(+)}(\tau_3)E_{S2}^{(+)}(\tau_2)|\Psi\rangle\right|^2 = \sum_{k_{S1}}|A_2(\tau_2,\tau_3)|^2, \tag{S19}$$

where $A_3(\tau)$ is often referred to as the three-photon amplitude or triphoton waveform. Notably, $A_2(\tau)$ also represents the three-photon amplitude, even though one subsystem is not detected in the experiment. It's essential to emphasize that both $A_3(\tau)$ and $A_2(\tau)$ are defined with reference to photon detections. By plugging Eq. (S13) into Eq. (S18), we attain:

$$A_3(\tau_1,\tau_2,\tau_3) = A_3 \sum_{k_{S1}}\sum_{k_{S2}}\sum_{k_{S3}} e^{-i(\omega_{S1}\tau_1+\omega_{S2}\tau_2+\omega_{S3}\tau_3)}F(k_{S1},k_{S2},k_{S3}), \tag{S20}$$

where all slowly varying terms and constants have been absorbed into $A_3$. Similarly, by substituting Eq. (S13) into Eq. (S19), we get:

$$A_2(\tau_2,\tau_3) = A_2 \sum_{k_{S2}} \sum_{k_{S3}} e^{-i(\omega_{S2}\tau_2+\omega_{S3}\tau_3)} F(k_{S1},k_{S2},k_{S3}), \tag{S21}$$

where again, all the slowly varying terms and constants have been grouped into $A_2$. Furthermore, to evaluate the Dirac $\delta$ function in $F$ (S14), we replace the summation over wavenumber with an angular frequency integral as usual,

$$\sum_{k_{Sj}} \rightarrow \frac{L}{2\pi}\int d\omega_{Sj}\frac{dk_{Sj}}{d\omega_{Sj}} = \frac{L}{2\pi}\int \frac{d\omega_{Sj}}{v_{Sj}}. \tag{S22}$$

Using Eqs. (S13) and (S22), the three-photon amplitude (S20) becomes

$$A_2(\tau_{21},\tau_{31}) = A_3 \int\int d\delta_2\, d\delta_3 \chi^{(5)}(\delta_2,\delta_3)\mathrm{sinc}\left[\frac{\Delta k(\delta_2,\delta_3)L}{2}\right] e^{-i\delta_2(\tau_{21}+L/2v_{S2})} e^{-i\delta_3(\tau_{31}+L/2v_{S3})}. \tag{S23}$$

The three-photon coincidence counting rate (S18) is simply modulus squared of $A_3(\tau_{21},\tau_{31})$, i.e., $R_3 = |A_3(\tau_{21},\tau_{31})|^2$.

From Eq. (S23), it is evident that the three-photon amplitude $A_3(\tau_{21},\tau_{31})$ is the convolution of the fifth-order nonlinear susceptibility $\chi^{(5)}(\delta_2,\delta_3)$ and the longitudinal phase-mismatch function $\Phi(\Delta kL)$. Physically, this implies that the triphoton temporal coherence is jointly determined by these two factors. As a consequence, we anticipate the appearance of two distinct regions in three-photon temporal correlation measurements, characterized by the damped Rabi oscillation regime dominated by $\chi^{(5)}$ and the group-delay regime dominated by $\Phi(\Delta kL)$. These regions have been explored in the experiment, and the recorded data are presented in Figs. 2–4 of the main text, as well as in Supplementary Figs. S11 and S12 (below). For qualitative comparison, Fig. S4 provides the corresponding theoretical simulations. It is apparent that both Figs. S4A and S4B exhibit the three-photon coincidence counts in the damped Rabi oscillation regime, while Fig. 4C displays the case in the group-delay region, qualitatively explaining the experimental observations in Figs. 2A, 3A, and 3D of the main text.

Similarly, we can demonstrate that the conditioned two~photon coincidence counting rate can be computed as:

$$R_2(\tau_{23}) = R_2 \int d\delta_3 \left| \int d\delta_2\, \chi^{(5)}(\delta_2,\delta_3)\mathrm{sinc}\left[\frac{\Delta k(\delta_2,\delta_3)L}{2}\right] e^{-i\delta_2(\tau_{23}+L/2v_{S2})} \right|^2, \tag{S24}$$

where $\tau_{23} = \tau_2 - \tau_3$ and $R_2$ is a grouped constant. As evident from Eq. (S24), $R_2(\tau_{23})$ is a function of $\tau_{23}$, indicating the presence of partial entanglement between the remaining $E_{S2}$ and $E_{S3}$ photons after tracing away the $E_{S1}$ photon. This unequivocally signifies the tripartite W-class property.

In Eq. (S24), the second integral inside the modulus squared is a convolution between $\chi^{(5)}$ and $\Phi(\Delta kL)$. Similarly, the functional profile of $R_2(\tau_{23})$ is in general determined by both factors. However, if one of these factors predominates, $R_2(\tau_{23})$ will showcase two distinctive scenarios: the damped Rabi oscillation regime and the group-delay regime. Other configurations for conditional two~photon coincidence counts can be calculated using the same logic. Here, we refrain from reiterating those computations and leave them as an exercise for the reader. In Figs. 2C and 2D, as well as Figs. 3B, 3C, 3E, and 3F in the main text, we present examples of such measured conditional two~photon coincidence counts. For qualitative comparison, Fig. S5 provides the corresponding theoretical simulations. It is evident that our theoretical framework aligns qualitatively with the experimental results. However, when comparing qualitative calculations with actual experimental data, significant differences become apparent. Wen and his colleagues have recently advanced in precise calculations of linear and nonlinear optical responses using the harmonic expansion method he developed. Initial calculations indicate promising agreement between theory and experiment. Further verification is underway, and the detailed results will be published in a separate publication.

Triphoton W State Entangled in Other Degrees of Freedom

While the primary focus of this study revolves around time-energy-entangled W triphotons, it is important to acknowledge that these W-class triphotons can also be readily entangled in other degrees of freedom, encompassing space-momentum, polarization, and orbital angular momentum. In other words, our work uniquely provides a dependable genuine W-class triphoton source, capable of generating a range of three-photon W states entangled across diverse degrees of freedom without involving additional interferometry setups or postselection. For example, our triphoton source can effortlessly yield triphotons entangled in space or momentum due to phase matching. Our source can also directly produce polarization-based W triphotons, without necessitating an interferometer, by inputting three linearly polarized cw lasers. The heightened SSWM process facilitated by atomic coherence enables the exploration of diverse forms of three-photon entanglement based on different degrees of freedom. This would be challenging or even unattainable using previously proposed schemes or methods.

Furthermore, our system exhibits the capability to generate triphoton hyperentangled states, entangling more than one degree of freedom of light. This introduces a significant technical challenge for any multiphoton generation platform reported thus far. While the system layouts and theoretical calculations concerning these triphoton entangled states are beyond the scope of this work, they will be elaborated upon in the forthcoming discussions.

Significantly, triphotons entangled in distinct degrees of freedom offer unique opportunities for quantum technological applications. For instance, the W-type triphotons endowed with spatial correlations (*52-54*) can be harnessed for quantum imaging and remoting sensing, enabling sub-Rayleigh superresolution that is both beyond the capabilities of biphotons (or entangled photon pairs) and classical light. This solidifies the fundamentally quantum nature of these phenomena and their potential to redefine quantum technologies.

Beyond the primary focus on the continuous-mode scenario explored in this study, our system seamlessly extends its capabilities to encompass the continuous variable (CV) regime. Within this framework, the generation of non-Gaussian tripartite states becomes a tangible achievement, facilitating their utilization across a spectrum of CV-based quantum information and computing protocols (*55*). This underscored adaptability and versatility inherent in our triphoton source stand as distinguishing features, setting it apart from many preceding methodologies overviewed in the main text.

Addressing Misconceptions: Clearing Up Common Misunderstandings

In what follows, we would like to clarify some misconceptions prevalent in studies related to multiphoton generation. Through careful examination of the existing literature, we are aware of several prevalent misunderstandings in the realm of multiphoton entanglement generation:

- *Equating multiphoton source with specific multiphoton state*. A prevalent misconception arises when the community conflates an “entangled multiphoton source” with “the realization of a specific entangled state.” It's crucial to discern the fundamental distinction between these two concepts. The former encompasses the latter comprehensively, while the latter represents only a singular instance. Our work’s significance lies in introducing a reliable genuine W-class triphoton source, capable of generating diverse three-photon W states entangled across various degrees of freedom—eliminating the need for additional interferometry and postselection.

Although we demonstrated time-energy triphoton entanglement, our source effortlessly produces triphotons entangled in space or momentum due to phase matching. This starkly contrasts with most prior multiphoton state demonstrations, which only achieve detection potential without acting as dependable multiphoton sources. Our approach, in contrast, ensures exclusive production of desired states due to the unique phase matching, guaranteeing confident, high-purity, and high-fidelity triphoton generation.

- *Comparing incompatible classes*. Recognizing the essential incongruity between the GHZ and W classes is paramount. This divergence underscores the importance of contextualizing the superiority of one class over the other within specific problems or applications. Without this contextual framework, any comparison lacks substantive relevance, rendering it incapable of enriching our understanding of multipartite entanglement. Furthermore, this inherent incompatibility leads to an intriguing consequence: any endeavor to transform a given class into its opposite counterpart demands the incorporation of supplementary interferometric setups and postselection measurements. Failing to do so renders such conversions unattainable.
- *Multiphoton production utilizing cascaded SPDCs/SFWMs*. the utilization of cascaded SPDCs or SFWMs for generating time-energy triphotons has demonstrated constrained dependability and suboptimal fidelity. This issue stems from the necessity of preserving over thousands of single photons resulting from the initial SPDC or SFWM process, awaiting the emergence of a singular pair from the subsequent process. As a consequence, in the absence of sophisticated detection systems, ensuring consistent production of a solitary triphoton entity remains elusive. This inherent limitation renders the feasibility of these methodologies ineffectual for establishing a dependable and authentic triphoton source.
- *Comprehending multiphoton entanglement with biphoton knowledge*. While our physics research typically begins by comprehending low-dimensional and simple scenarios before attempting to generalize to high-dimensional and complex situations, it is crucial to be mindful of the limitations of this approach when studying multi-photon entanglement. One significant challenge, as repeatedly emphasized, arises from the fact that, unlike the biphoton case, the triphoton scenario involves two incompatible classes. Understanding these properties goes beyond the knowledge derived from the biphoton context.

It also comes to our attention that even though multiphoton polarization-based GHZ state (including the four-photon case) can be derived from SPDC or SFWM processes—given that these processes generate photons in pair—one must recognize that these multiphoton outputs arise from higher-order perturbation terms. To effectively detect such photon states, the construction of sophisticated detection systems becomes imperative in order to mitigate accidental counts stemming from dual photon pairs. Without effectively mitigating these prevalent photon trigger events originating from lower perturbations, the viability of establishing a reliable source using this scheme remains unattainable.

All in all, despite the multitude of protocols proposed over the last two decades for generating multiphoton entangled states, as comprehensively discussed in the main text, our perspective suggests that none of these protocols have matured into dependable multiphoton sources. This sentiment is rooted in the presence of inherent limitations and external challenges within these methodologies. Conversely, our devised scheme emerges as the most promising candidate to date for realizing a genuinely practical W-class triphoton source, bringing us notably closer to achieving this elusive goal.

## II. Further Insights into Experimental Measurements and Data Processing

In the subsequent subsections, we would like to delve into the experimental measurements and data processing with greater depth. Additionally, we will present an extended collection of experimental data on triphoton coincidences, offering further evidence that the suggested SSWM process within coherent atomic ensembles efficiently produces genuine triphotons of exceptional quality and reliability. Notably, these supplementary findings, combined with the data presented in the main text, provide a comprehensive illustration of the source's versatility. This versatility holds the potential to unlock novel technological advancements that are currently beyond the reach of existing photon resources.

Possible Biphoton Processes

As outlined in the Methods section, a significant source of accidental coincidence noise in the three-photon correlation measurements mainly stem from the simultaneous occurrence of two pairs of biphotons, originating from distinct spontaneous four-wave mixing (SFWM) processes, detected by the single-photon detectors. Fortunately, these SFWMs exhibit differing phase matching conditions, deviating from the one pertinent to the SSWM process. Furthermore, the biphotons resulting from these SFWMs possess distinct central frequencies in contrast to those of the desired triphotons.

By meticulous manipulation of the phase matching conditions and the implementation of narrowband filters, the false trigger events from these biphotons can be effectively eliminated from the authentic triphoton coincidence counts. For a visual representation of these biphoton generation scenarios, Fig. S6 provides a schematic depiction of all possible SFWM processes. Leveraging the level structure, seven such SFWM processes have been identified and visually presented in Figs. S6B–H. It's worth noting that the biphotons originating from these SFWMs constitute the primary source of accidental coincidences impacting the actual measurements. In the Methods section, we have extensively expounded upon the potential combinations of these SFWM processes that could lead to error-triggering events.

While it is theoretically possible to generate entangled quadraphotons through higher-order nonlinear wave mixing processes, the likelihood of their emission remains considerably low. Consequently, they do not pose a significant noise source for triphoton detection. Given this context, we will refrain from delving further into the discussion of entangled quadraphotons in this context.

Coincidence Counts obtained by Background Accidental Subtraction

Figures 2 and 3 in the main text showcase the recorded data alongside background accidental counts. In the corresponding Figs. S7 and S8, we present the same measured data after background accidental counts have been subtracted. A comparison between Figs. 2 and 3 and Fig. S7 and S8 underscores that the crucial characteristics remain well-preserved in both instances.

In Figs. S7C, S7D, S8B, S8C, and S8F, we have incorporated green and red dashed lines based on the measured data to highlight the oscillation periods referenced in the main text. By juxtaposing Figs. S7C, S7D, S8B, S8C, S8E, and S8F with Figs. S4A1–C2, we acknowledge that our qualitative optical response model can only furnish a qualitative interpretation of the experimental outcomes. Nonetheless, this approach effectively reveals fundamental features within the measurements.

For a more comprehensive grasp of both conditional two~photon coincidences and conditional three-photon coincidences, we have extended our analysis beyond Figs. 2A, 3A, and 3D in the main text. By carefully removing the corresponding background accidental counts and exploring varied scenarios, we gain deeper insights. Figure S9 serves as an illustrative example of this processed experimental data, meticulously organized to adhere to specific conditions. Within these figures, it becomes evident that the coherence length of the residual temporal correlation for the two~photon scenario is not fixed; rather, it varies contingent upon the specific measurement conditions. This variability similarly extends to the coherence length of the conditional three-photon temporal correlation. Importantly, these dynamic features were not discernible in prior demonstrations. From an alternative perspective, this observation also substantiates the adaptability and adjustability inherent in the generated three-photon state—a crucial attribute for its diverse range of applications.

As a W state, the outcome of tracing out the $E_{S1}$-photons raises an intriguing question. Figures S10A-C respectively report the resulting conditioned two~photon coincidence counts between the remaining $E_{S2}$ and $E_{S3}$ photons for the cases shown in Figs. 2A, 3A, and 3D of the main text. Upon observation, we find that these profiles starkly differ from those illustrated in Figs. 2C, 2D, 3B, and 2F of the main text, as well as Figs. S7C, S8B, S8E, S8F, S9A1-C1, SBA2-C2, S12B, S12C, S13B, and S13C within the SI. The profiles manifested in Fig. S10 below are indeed anticipated, as the $E_{S1}$-photons do not experience the slow-light effect. As a result, the residual temporal correlations between the remaining $E_{S2}$ and $E_{S3}$ photons assume a nearly symmetrical distribution around the origin of time ($\tau_{32} = 0$).

Procedure for Reconstructing Triphoton Coincidence Counts

Unlike standard two-photon correlation measurements, it's worth noting that a commercially available generic three-photon coincidence circuit is absent in the current market landscape. Consequently, research groups are tasked with constructing their own dedicated three-photon coincidence circuits. As depicted in Fig. S11, we establish a detection system based on two-photon coincidence circuits. Specifically, within a predetermined three-photon correlation time window, we reconstruct three individual single-photon trigger events from $SPCM_1$, $SPCM_2$, and $SPCM_3$. This reconstruction is achieved through the simultaneous detection of two pairs of two-photon coincidence counts, namely $\{E_{S1}, E_{S2}\}$ and $\{E_{S1}, E_{S3}\}$, facilitated by an additional diagnostic $SPCM_D$.

In practical experimentation, for each recorded three-photon coincidence count, the $E_{S1}$-photon click serves as a shared start trigger, initiating two electronic pulses I1 from $SPCM_1$. One of these pulses is subjected to a 150-ns delay, as depicted in Fig. S11A. Concurrently, the detections of the $E_{S2}$-photon and $E_{S3}$-photon serve as the stop trigger. Here, the electronic pulse $I_3$ from $SPCM_3$ is delayed by 150 ns relative to the electronic pulse $I_2$ from $SPCM_2$. With these intricate setups, the measurement of $E_{S1}$ and $E_{S2}$ photons is conducted first as a function of $\tau_{21}$, followed by the recording of $E_{S1}$ and $E_{S3}$ photons after a 150 ns interval, captured as a function of $\tau_{31}$. This methodology allows for the capture of three-photon temporal correlations within the context of coincidence counting measurements.

To illustrate the functioning of each two-photon coincidence counting component, Fig. S11B-D showcases a representative set of experimental data collected over a span of 5 minutes, employing a time bin width of 0.25 ns for each SPCM. It is evident that the joint detection of $E_{S1}$ and $E_{S2}$ photons elicits a two-photon temporal correlation, varying according to the relative time difference $\tau_{21}$ between the clicks of the involved single-photon detectors, $SPCM_1$ and $SPCM_2$ (Fig. S11B). Similarly, the joint detection of $E_{S1}$ and $E_{S3}$ photons unveils a residual temporal correlation, contingent upon the relative triggering time difference $\tau_{31}$ between the clicks of the engaged single-photon detectors, $SPCM_1$ and $SPCM_3$ (Fig. S11C). As the diagnostic single-photon detector $SPCM_D$ is triggered by artificial electronic signals, coincident counting between $E_{S3}$ photons and these artificial diagnose signals yields no exact temporal correlation, as demonstrated in Fig. S11D.

Experimentally, capturing authentic triphotons through detection hinges critically on optimizing the phase-matching conditions of the SSWM process. This optimization is achieved by controlling the wavelengths and injection angles of the three input optical driving beams, alongside the directions of triphoton collection. Beyond these arrangements, an additional layer of assurance is established to confirm that the detected triphotons originate exclusively from the intended SSWM process.

This assurance is accomplished by implementing coincident counting detection. Here, the $E_{S3}$ photons are jointly measured with artificially introduced diagnostic signals originating from $SPCM_D$. This joint measurement transpires concurrently with the combined detection of $E_{S1}$ and $E_{S2}$ photons. Utilizing the same reconstruction method outlined earlier, we obtain merely a few accidental coincidences per minute when employing the two-photon coincidences $\{E_{S1}, E_{S2}\}$ and $\{E_{S3}, E_D\}$ to construct the three-photon histogram. This outcome underscores the absence of any authentic quantum correlation within any two pairs of unrelated two-photon coincidences.

Supplementary Experimental Data

In the experimental domain, we conducted a series of three-photon coincidence counting measurements while varying system parameters. In addition to the data depicted in Figs. 2 and 3 in the main text, we present an additional set of measured data. Illustrated in Fig. S12, we accumulated three-photon coincidence trigger events over the course of 1 hour, utilizing a time bin width of 2.0 ns for each SPCM. Most experimental parameters remain consistent with those detailed in Fig. 2A of the main text, except for $P_2$ (7 mW), $P_3$ (7 mW), and $OD$ (45.7).

From the recorded data, it emerges that the triphoton production rate is 100 ± 11 per minute, accompanied by background accidentals of 8 ± 3.1 per minute. Notably, even in this scenario, the triphoton temporal correlation remains within the group-delay regime. This is substantiated by evaluating the conditional two~photon correlations, achieved by tracing away one photon from each triphoton. Fig. S12B and S12C present these conditional two~photon coincidence counts. It is evident that the previously observed Rabi oscillations almost diminish in these two figures.

Illustrated within Fig. S13, we present an additional series of measurements within the group-delay region. A direct comparison with Fig. S12 reveals a significant reduction in the amplitude of the small oscillations observed in the preceding figures.

## III. Summary of Diverse Mechanisms for Multiphoton Generation

In this section, we have consolidated the primary experimental demonstrations showcasing the generation of entangled three-photon and multiphoton states, which have been documented up to this point. We've compiled their essential parameters and resulting optical properties in TABLE I, providing a convenient point of reference. It is important to acknowledge that our intention is not to list every single experimental report within this compilation. Nonetheless, the reports included here serve as somewhat representative examples of the broader landscape.

## IV. Extended Discussion on the Reported Triphoton Source

It is illuminating to investigate the feasibility of the reported triphoton source in generating GHZ-type triphotons entangled in time-energy (and other degrees of freedom) (*56*, *57*). To our current understanding, the literature lacks any single proposal for the direct creation of continuous-mode time-energy-entangled GHZ triphotons. This absence stems from the requirement that, in order to establish such a three-photon GHZ state, two of the photons must be degenerate in all degrees of freedom (*54*, *56*, *57*).

**TABLE I**

| Class | State | Counts per hour | Reference |
|---|---|---|---|
| Triphoton Source | time-energy (cascaded SPDCs) | 6.2 | *Nature* **466**, 601 (2010) |
| | time-energy (cascaded SPDCs) | 7 | *Nat. Phys.* **9**, 19 (2013) |
| | polarization GHZ (cascaded SPDCs) | 744 | *Nat. Photon.* **8**, 801 (2014) |
| | **W for various degrees of freedom (SSWM)** | **7500** | ***This work*** |
| Specific Multiphoton State Generation | 3-photon polarization GHZ (SPDC) | 24 | *Phys. Rev. Lett.* **82**, 1345 (1999) |
| | 4-photon polarization GHZ (SPDC) | 69 | *Phys. Rev. Lett.* **86**, 4435 (2001) |
| | 4-photon polarization GHZ (SPDC) | 300 | *Phys. Rev. Lett.* **90**, 200403 (2003) |
| | 4-photon polarization (SPDC) | 175 | *Phys. Rev. Lett.* **92**, 107901 (2004) |
| | 5-photon polarization GHZ (SPDC) | 10 | *Nature* **430**, 54-58 (2004) |
| | 3-photon polarization W (SPDC) | 5220 | *Phys. Rev. Lett.* **95**, 150404 (2005) |
| | 4-photon polarization Dicke (SPDC) | 3600 | *Phys. Rev. Lett.* **98**, 063604 (2007) |
| | 3-photon discrete-energy W (SFWM) | 75 | *Phys. Rev. Lett.* **123**, 070508 (2019) |
| | 4-photon polarization GHZ (SFWM) | 2088 | *Adv. Quantum Tech.* **4**, 2000152 (2021) |
| | 4-photon polarization GHZ (SFWM) | 6084 | *Appl. Phys. Lett.* **120**, 024001 (2022) |

With regard to our proprietary triphoton source, you might be intrigued by the possible outcome achieved through the arrangement of two of these photons into a degenerate state. Could such an arrangement potentially yield a GHZ state? In theory, such a scenario is indeed plausible. However, from a practical perspective, the execution of an experiment of this nature would present substantial challenges. Moreover, considering an alternative perspective, the development of a scheme for the direct generation of continuous-mode triphoton and multi-photon states entangled in time-energy and space-momentum domains still necessitates additional in-depth research efforts.

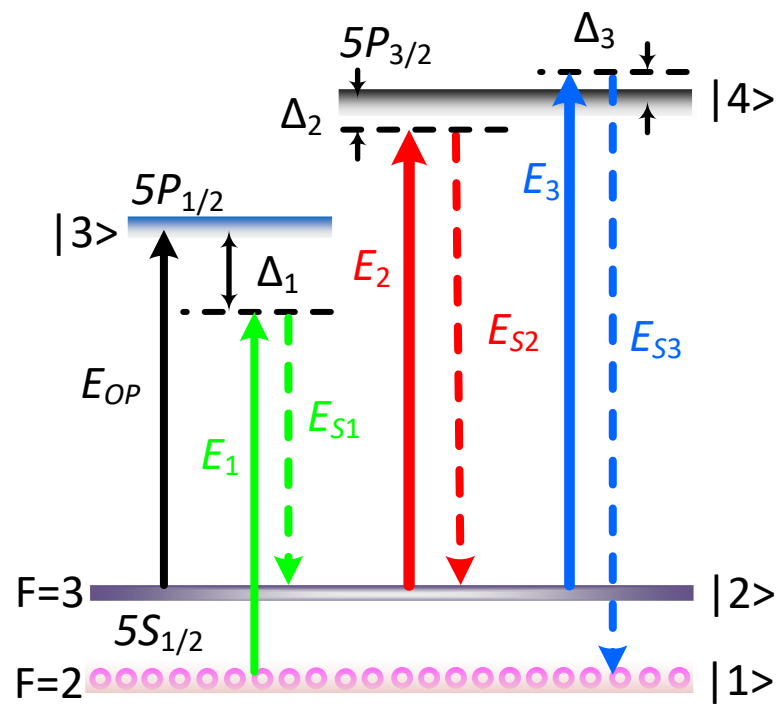

**Fig. S1.**

Energy-level diagram of hot [85]Rb atoms illustrating direct time-energy-entangled W-class triphoton generation. This four-level triple-Λ-type atomic configuration features two ground states $|1\rangle$ and $|2\rangle$, as well as two excited states $|3\rangle$ and $4\rangle$. Initial atomic population is established in state $|1\rangle$. To prevent residual atomic population in $|2\rangle$, an additional resonant optical pumping beam $E_{OP}$ is introduced for the atomic transition $|2\rangle \leftrightarrow |3\rangle$. A weak cw pump laser $E_1$ is directed towards $|1\rangle \rightarrow |3\rangle$ with a large, fixed red frequency detuning $\Delta_1$. Meanwhile, another two strong cw control fields, $E_2$ and $E_3$, are concurrently applied to the same atomic transition $|2\rangle \rightarrow |4\rangle$, but with different frequency detunings $\Delta_2$ and $\Delta_3$. By adhering to the required phase-matching conditions, the spontaneous six-wave mixing (SSWM) process is facilitated, enabling the direct and efficient emission of continuous-mode time-energy-entangled W-type triphotons—$E_{S1}$, $E_{S2}$ and $E_{S3}$—from their respective atomic transitions. This emission process is visually depicted in the diagram.

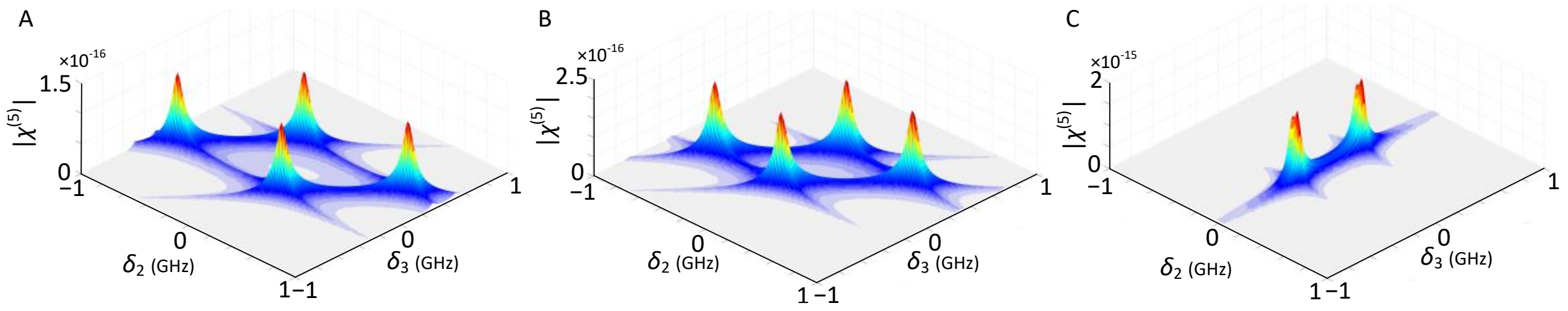

**Fig. S2.**

Exemplary visualization of the fifth-order nonlinear susceptibility $\chi^{(5)}$ across different parameter configurations. (**A**) $\chi^{(5)}$ corresponding to **Fig. 2A** of the main text, utilizing the following simulation parameters: $\Gamma_{31} = \Gamma_{41} = 2\pi \times 6$ MHz, $\Gamma_{11} = \Gamma_{22} = 0.4 \times \Gamma_{41}$, $\Gamma_{21} = 0.2 \times \Gamma_{41}$, $\Delta_1 = -2\text{GHz}$, $\Delta_2 = -150\text{MHz}$, $\Delta_3 = 50\text{MHz}$, $OD = 4.6$, $\Omega_1 = 300$ MHz, $\Omega_2 = 870$ MHz, and $\Omega_3 = 533$ MHz. Input laser powers are set at $P_1 = 4$ mW, $P_2 = 40$ mW, and $P_3 = 15$ mW. (**B**) $\chi^{(5)}$ corresponding to **Fig. 3A** of the main text, utilizing the same simulation parameters as (**A**), with the exception of $\Omega_2 = 533$ MHz and input power $P_2 = 15$ mW. (**C**) $\chi^{(5)}$ corresponding to **Fig. 3D** of the main text, employing the same simulation parameters as (**B**), except for $OD = 45.7$.

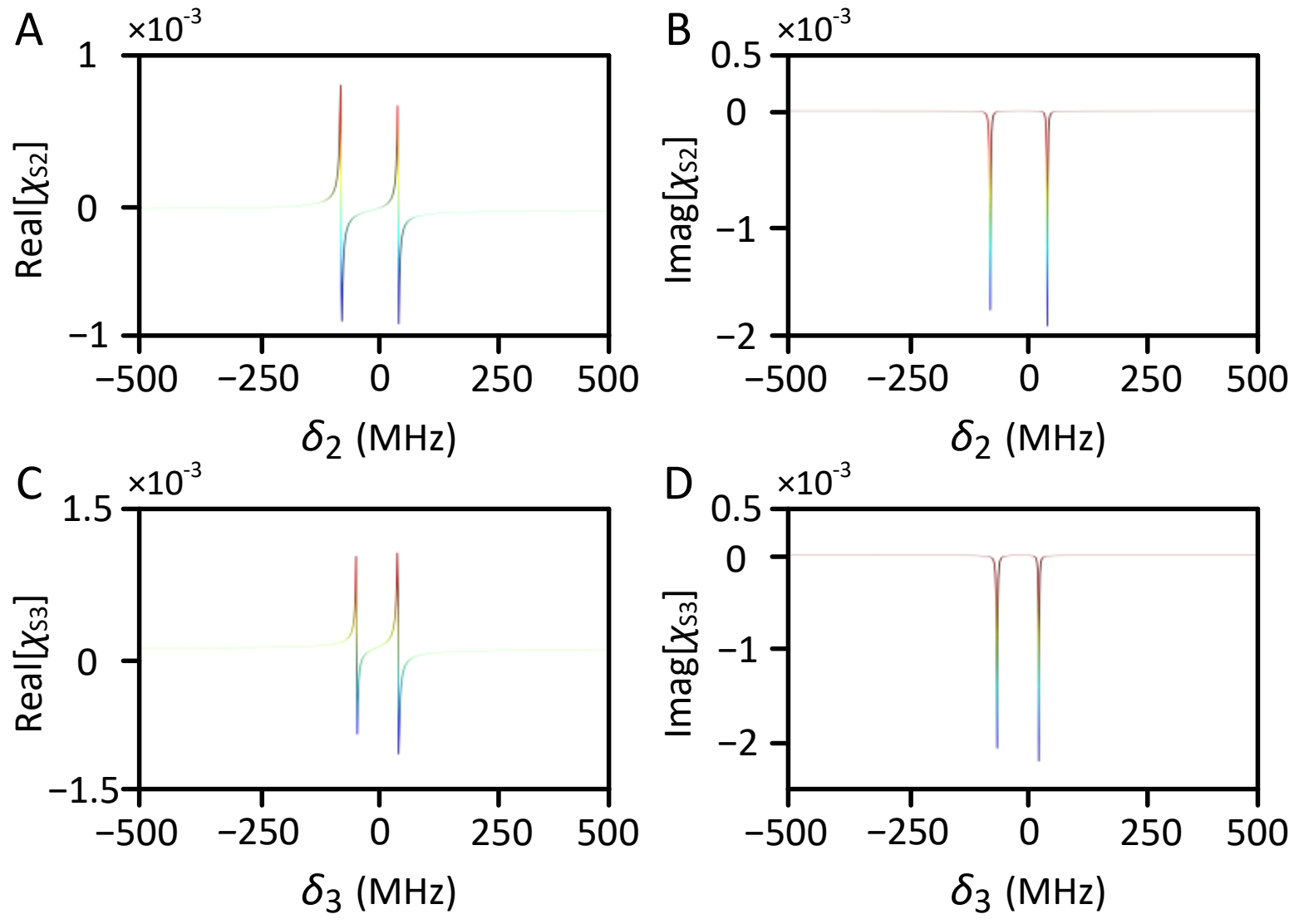

**Fig. S3.**

Representative illustrations of linear susceptibilities $\chi_{S2}$ and $\chi_{S3}$. The parameters involved remain consistent with those employed in **Fig. S2B**. (**A & B**) Display of the real and imaginary parts of $\chi_{S2}$. (**C & D**) Depiction of the real and imaging components of $\chi_{S3}$.

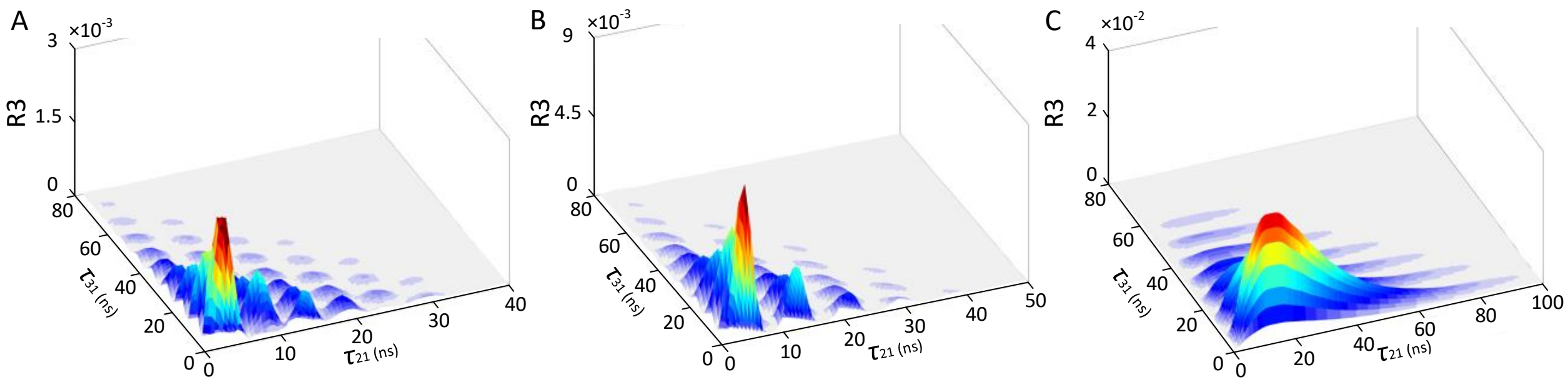

**Fig. S4.**

Theoretical simulations of triphoton coincidence counting rates $R_3$. (**A**) $R_3$ associated with **Fig. 2A** in the main text, employing identical parameters to those featured in **Fig. S2A**. (**B**) $R_3$ corresponding to **Fig. 3A** in the main text, utilizing the same parameters as those in **Fig. S2B**. (**C**) $R_3$ related to **Fig. 3D** in the main text, using the same parameters as those in **Fig. S2C**.

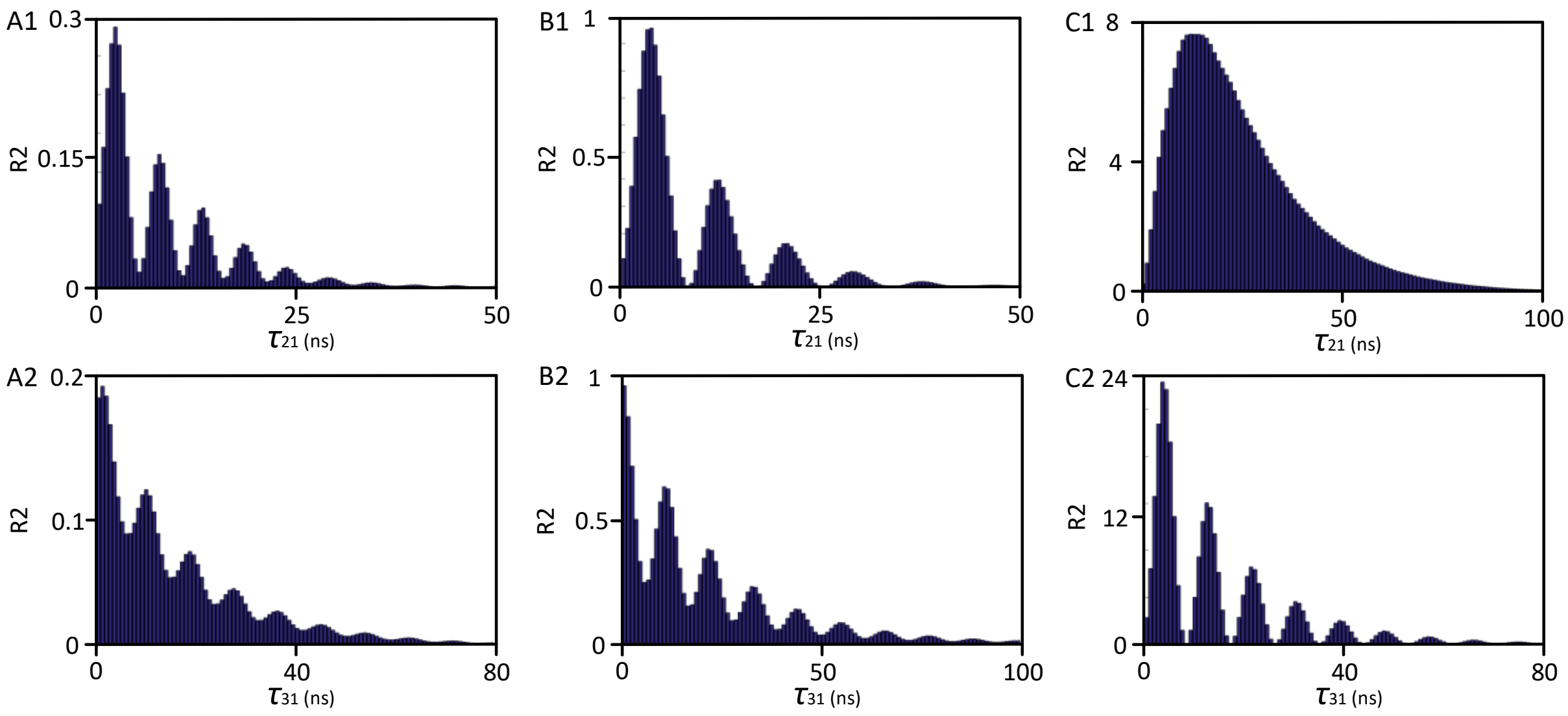

**Fig. S5.**

Theoretical simulations of conditional two~photon coincidence counting rates $R_2$ for Fig. S4. (**A1**) $R_2$ achieved by tracing away the $E_{S3}$-photons in **Fig. S4A**. (**A2**) $R_2$ attained by tracing away the $E_{S2}$-photons in **Fig. S4A**. (**B1**) $R_2$ acquired by tracing away the $E_{S3}$-photons in **Fig. S4B**. (**B2**) $R_2$ acquired by tracing away the $E_{S2}$-photons in **Fig. S4B**. (**C1**) $R_2$ achieved by tracing away the $E_{S3}$-photons in **Fig. S4C**. (**C2**) $R_2$ attained by tracing away the $E_{S2}$-photons in **Fig. S4C**.

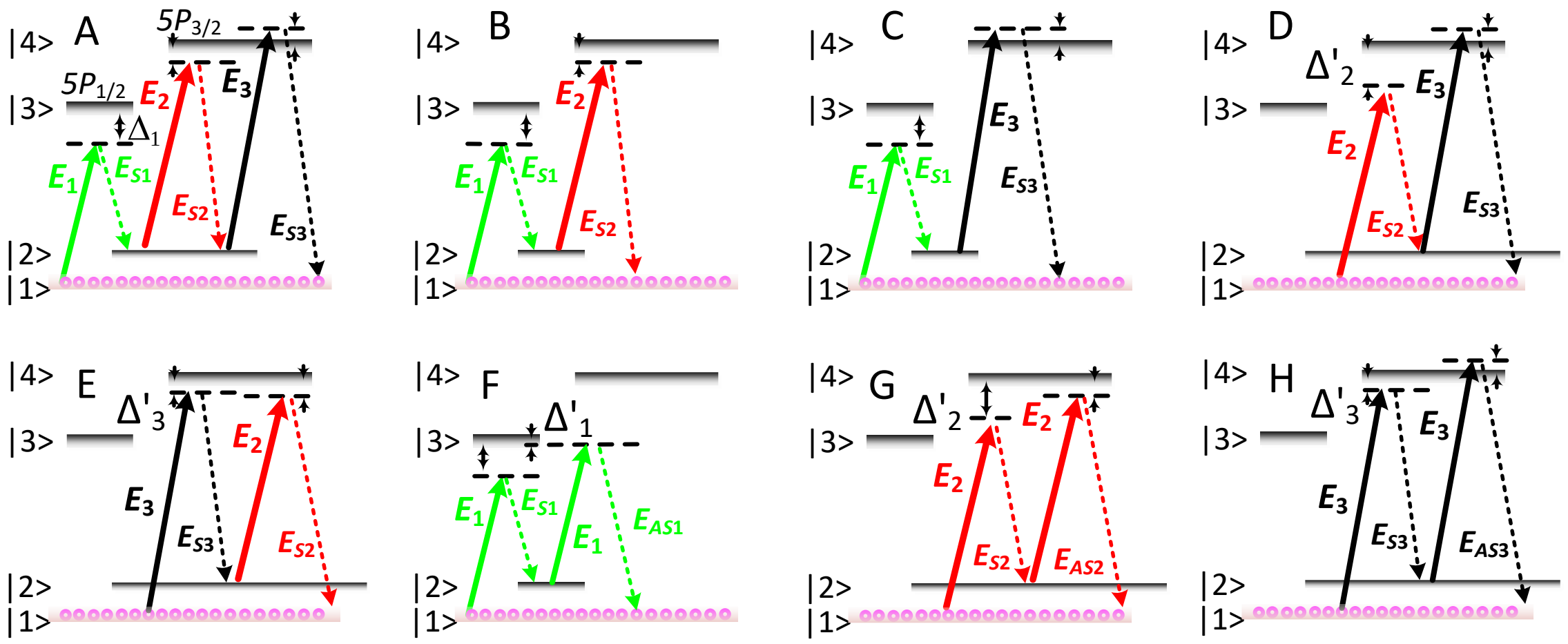

**Fig. S6.**

Seven potential SFWM processes leading to accidental coincidences in three-photon coincidence counting measurement. (**A**) Illustration of the atomic energy-level structure governing triphoton generation. (**B-H**) Seven distinct possible SFWM processes, each outlining scenarios where emitted biphotons might inadvertently contribute to accidental coincidences within the three-photon coincidence counts that are measured.

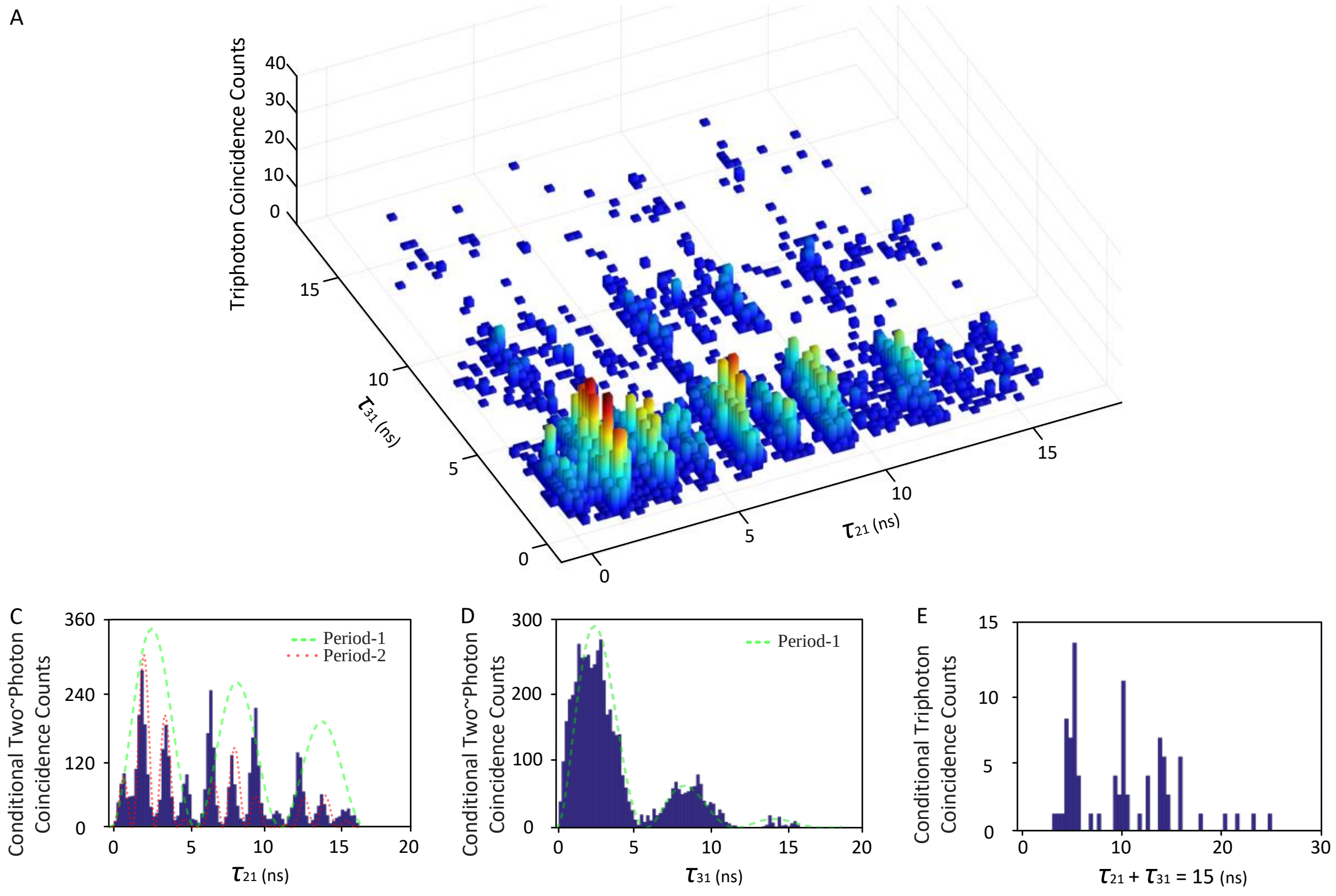

**Fig. S7.**

Triphoton coincidence counts and conditioned two~photon & three-photon coincidence counts from Fig. 2 (main text), after background accidental removal. In panels (**C & D**), periodic oscillations discussed in the main text are visualized using green and red dashed lines.

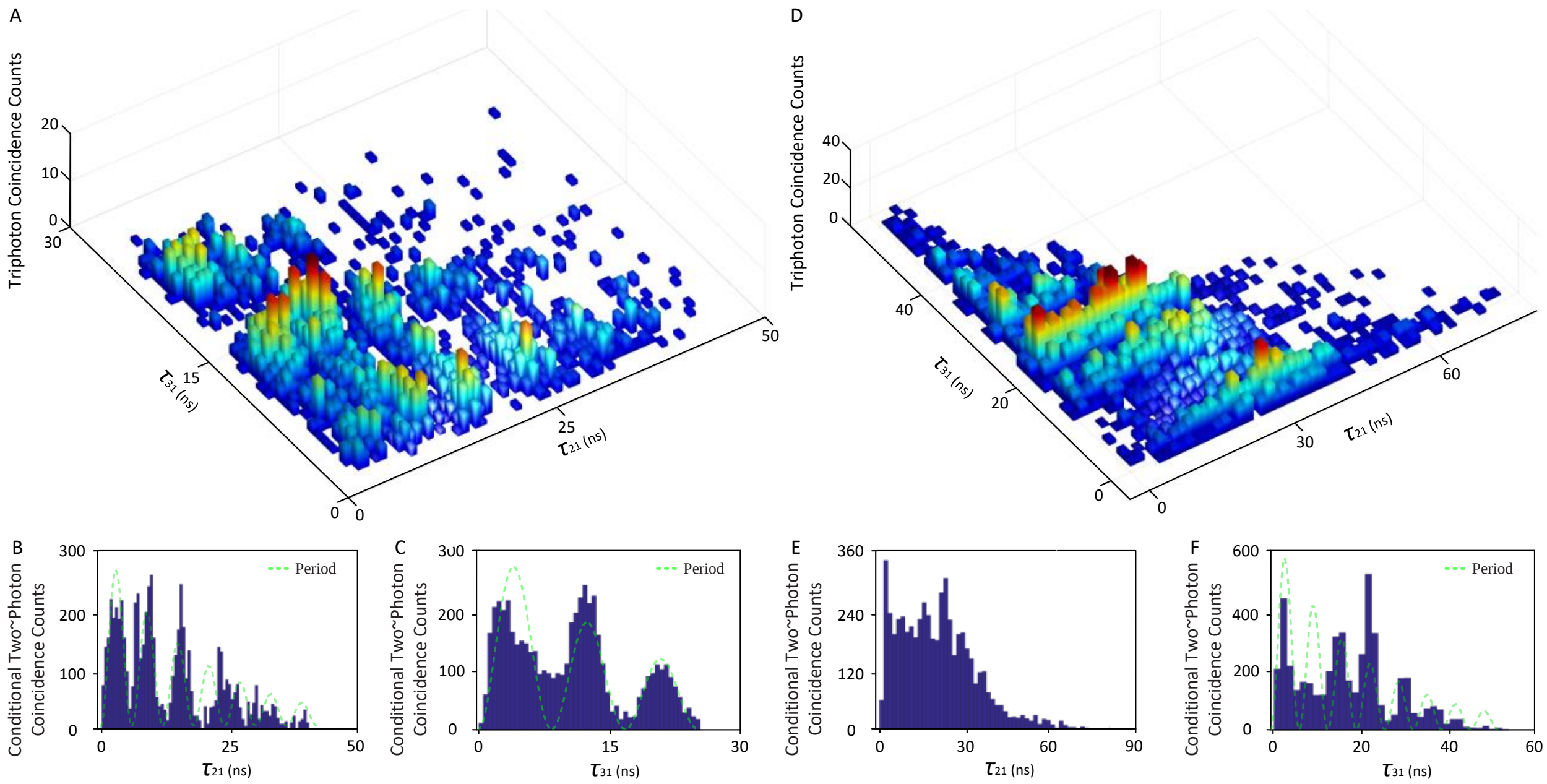

**Fig. S8.**

Triphoton coincidence counts and conditioned two~photon & three-photon coincidence counts from Fig. 3 (main text), after background accidental subtraction. In panels (**B, C & F**), the presence of periodic oscillations, as discussed in the main text, is visually highlighted through the use of green dashed lines.

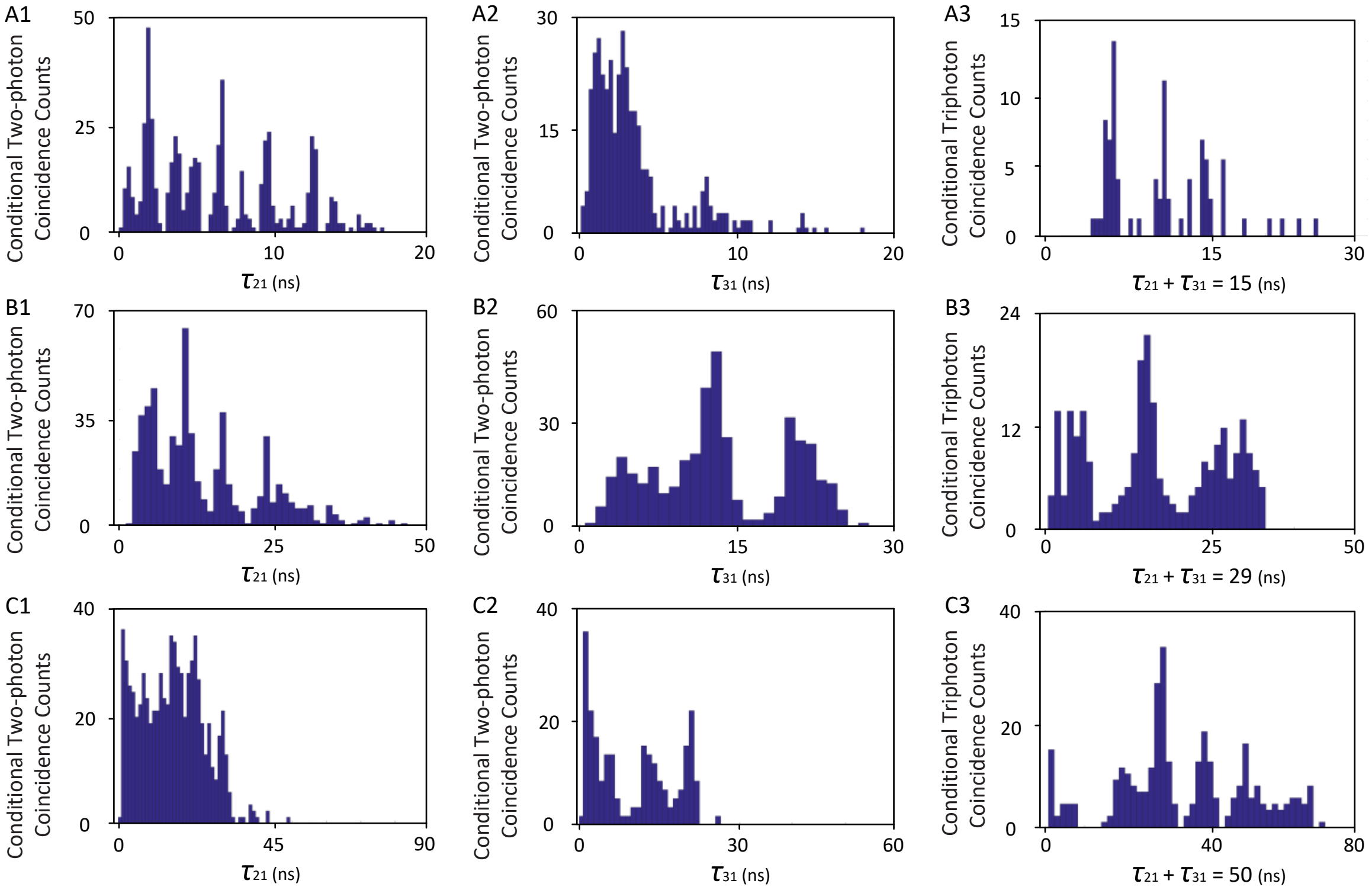

**Fig. S9.**

Conditional two~photon and triphoton coincidence counts, $R_2$ and $R_3$, with background accidental subtraction. Presented here are conditioned two~photon coincidence counts ($R_2$) and conditional three-photon coincidence counts ($R_3$) for the scenarios depicted in Figs. 2A, 3A, and 3D from the main text. Specifically, for Fig. 2A in the main text: (**A1**) $R_2(\tau_{21})$ with $\tau_{31} = 2.6$ ns for $R_3$; (**A2**) $R_2(\tau_{31})$ with $\tau_{21} = 2.0$ ns for $R_3$; (**A3**) $R_3(\tau_{21} + \tau_{31} = 15.0\text{ ns})$. For Fig. 3A in the main text: (**B1**) $R_2(\tau_{21})$ with $\tau_{31} = 13.0$ ns for $R_3$; (**B2**) $R_2(\tau_{31})$ with $\tau_{21} = 4.0$ ns for $R_3$; (**B3**) $R_3(\tau_{21} + \tau_{31} = 29.0\text{ ns})$ for $R_3$. For Fig. 3D in the main text: (**C1**) $R_2(\tau_{21})$ with $\tau_{31} = 21.0$ ns for $R_3$; (**C2**) $R_2(\tau_{31})$ with $\tau_{21} = 31.0$ ns for $R_3$; (**C3**) $R_3(\tau_{21} + \tau_{31} = 50.0\text{ ns})$ for $R_3$.

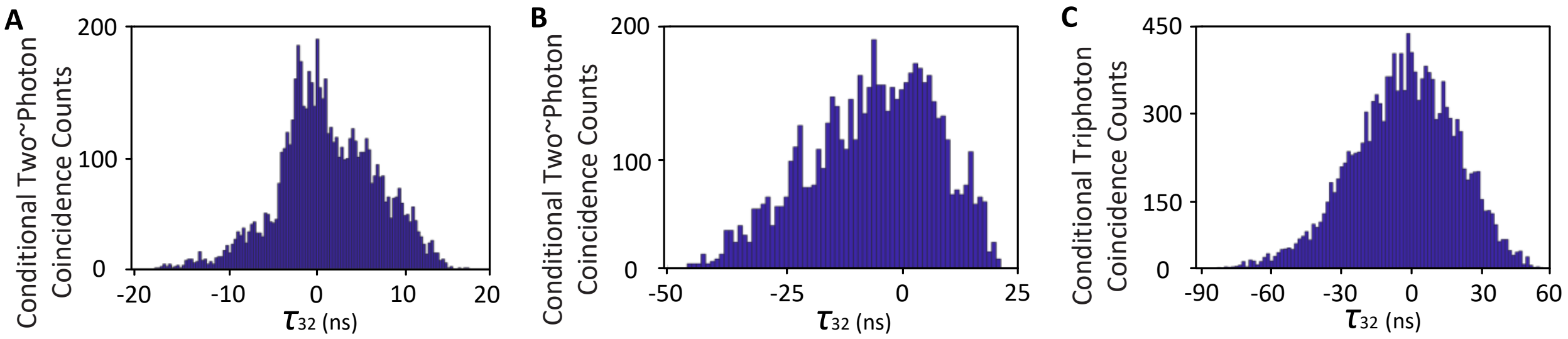

**Fig. S10.**

Conditional two~photon coincidence counts by tracing away $E_{S1}$-photons. Derived from the data schematic in Figs. 2A, 3A, and 3D of the main text, these plots depict conditional two~photon coincidence counts resulting from the removal of $E_{S1}$-photons.

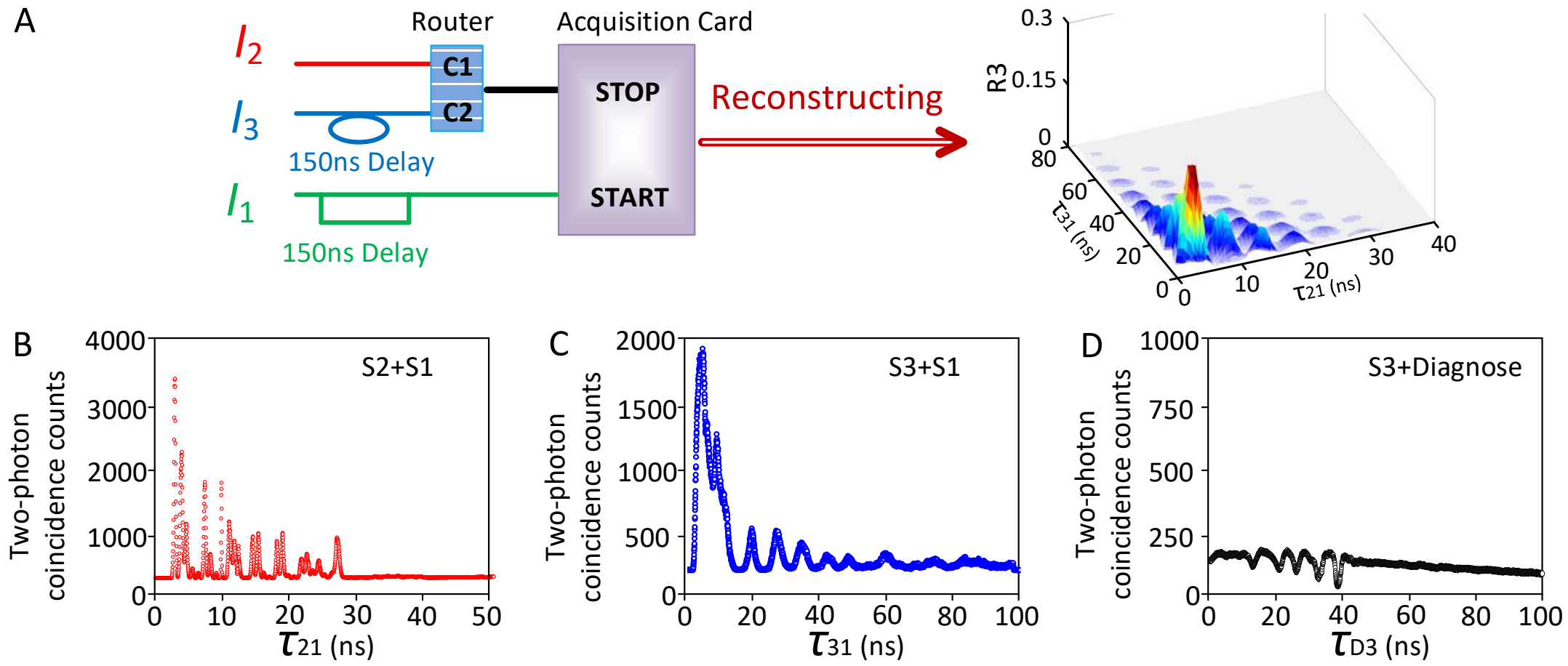

**Fig. S11.**

Three-photon detection system and coincidence counting reconstruction. (**A**) Illustrated schematic of our home-made detection system that facilitates the reconstruction of three-photon coincidence counting. As an illustrative example, panels (**B–D**) exhibit the recorded two-photon coincidence counts in one experiment, respectively, by $SPCM_1$ and $SPCM_2$, $SPCM_1$ and $SPCM_3$, and $SPCM_3$ and $SPCM_D$. These trigger events are plotted against the relative time differences ($\tau_2$, $\tau_3$ and $\tau_d$) between clicks of the two respetive single-photon detectors. The experimental data was accumulated over 5-minute period, utilizing a time bin width of 0.25 ns for each SPDCM. Additional parameters are set as follows: $P_1 = 4$ mW, $P_2 = 40$ mW, $P_3 = 15$ mW, $\Delta_1 = -2$GHz, $\Delta_2 = -150$MHz, $\Delta_3 = 50$MHz, $\Omega_1 = 300$ MHz, $\Omega_2 = 870$ MHz, and $\Omega_3 = 533$ MHz.

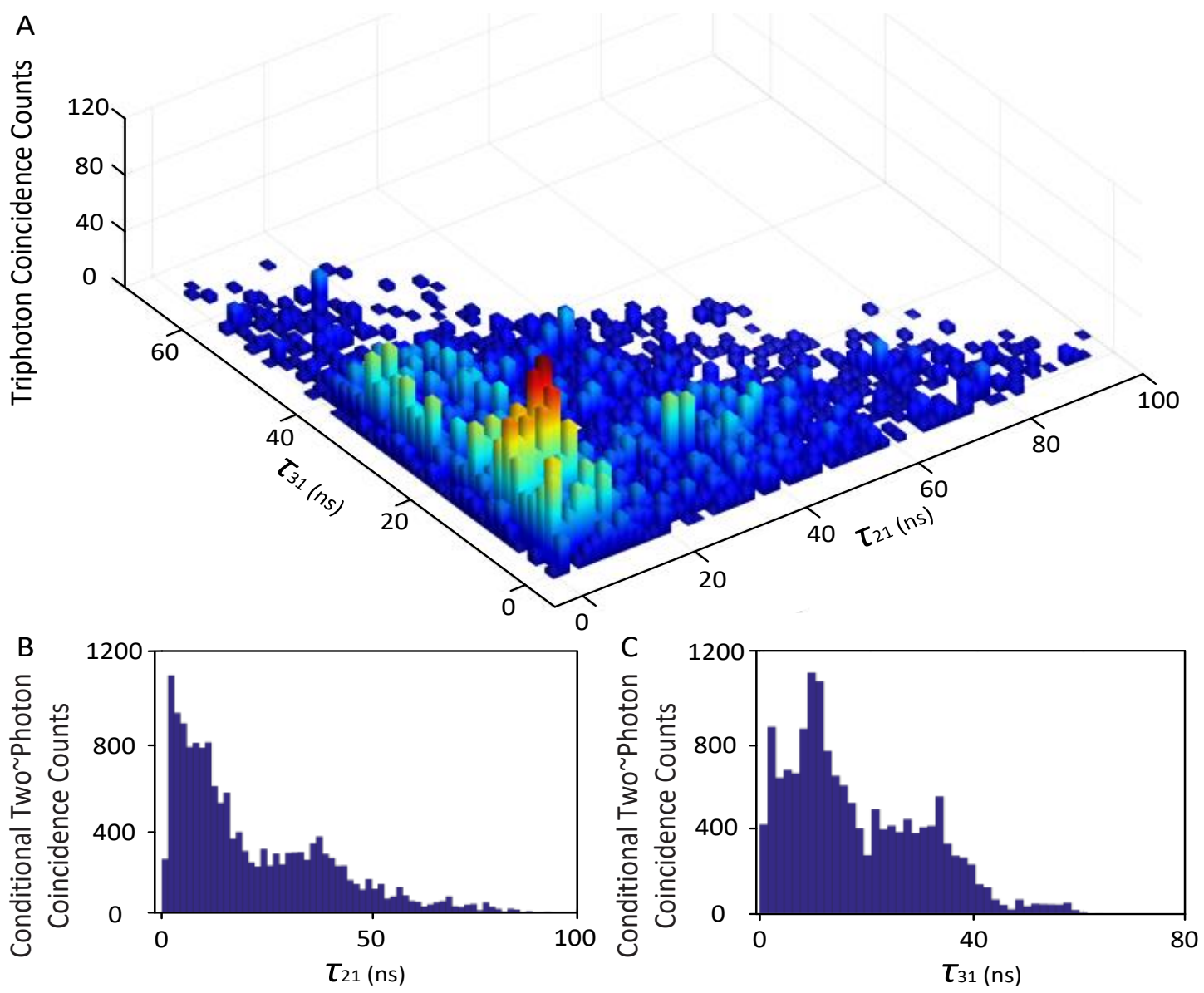

**Fig. S12.**

Triphoton temporal correlation in the group-delayed region. (**A**) Depiction of the histogram representing three-photon coincidence counts spanning 1 hour, utilizing a time-bin width of 2.0 ns for each single-photon detector. The triphoton generation rate amounts to 100 ± 11 per minute, accompanied by background accidental coincidences measured at 8 ± 3.1 per minute. (**B & C**) Conditional two~photon coincidence counts attained by tracing away the $E_{S3}$ or $E_{S2}$ photons from each respective three-photon joint trigger event displayed in panel (**A**). The experimental parameters match those of **Fig. 2A** in the main text, with exceptions being $P_2 = 7$ mW, $P_3 = 7$ mW, $\Omega_2 = 364$ MHz, $\Omega_3 = 364$ MHz, and $OD = 45.7$.

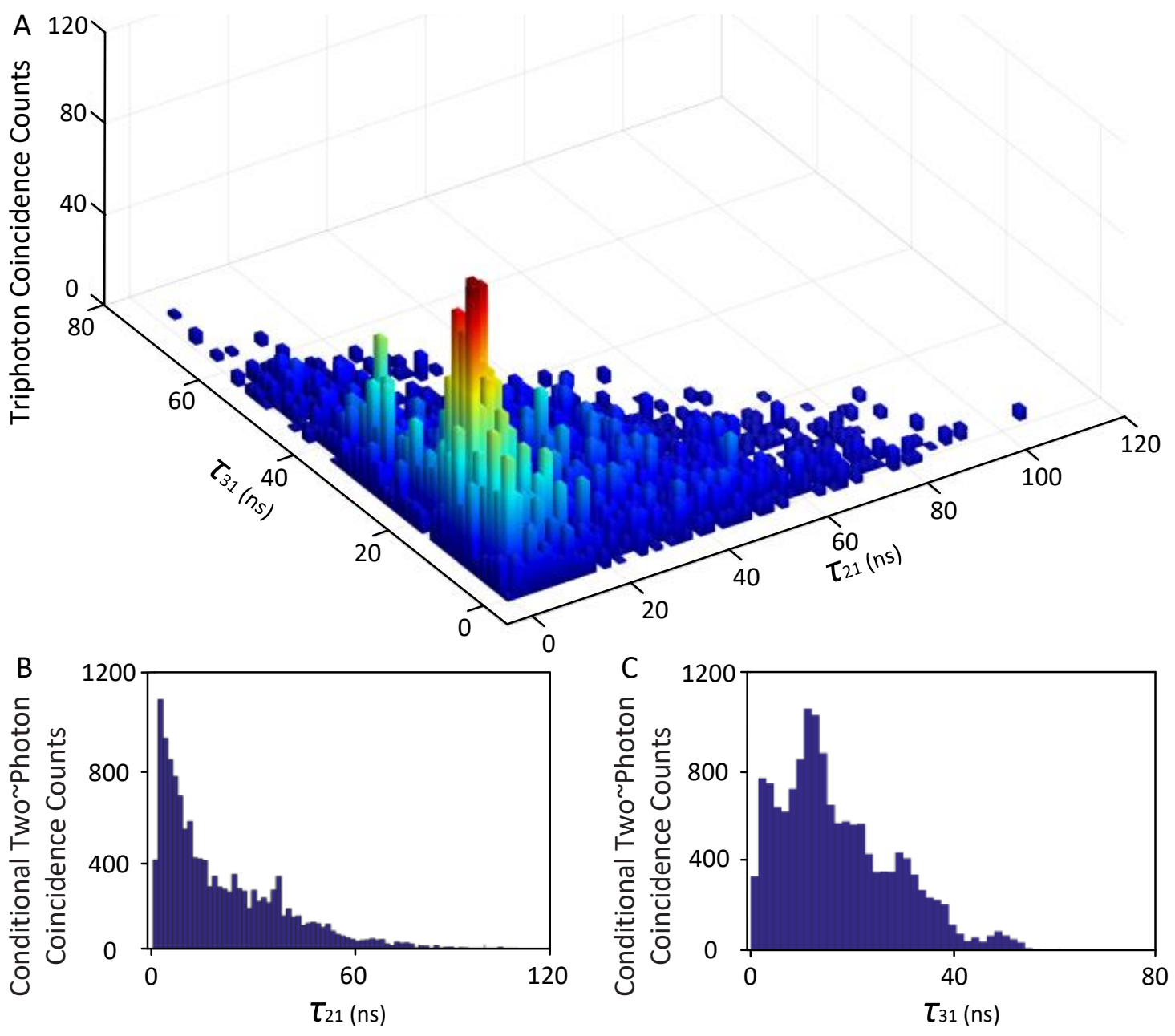

**Fig. S13.**

Triphoton temporal correlation in the group-delayed region. (**A**) Depiction of the histogram representing three-photon coincidence counts spanning 1.5 hours, employing a time-bin width of 1.5 ns for each individual single-photon detector. The triphoton generation rate is determined as 140 ± 15 per minute, with accompanying background accidental coincidences measured at 13 ± 3.4 per minute. (**B & C**) Conditional two~photon coincidence counts achieved through the elimination of the $E_{S3}$ or $E_{S2}$ photons from each respective three-photon joint trigger event presented in panel (**A**). The experimental parameters align with those of **Fig. 2A** in the main text, with alterations such as $P_2 = 6$ mW, $P_3 = 6$ mW, $\Omega_2 = 351$ MHz, $\Omega_3 = 351$ MHz, and $OD = 45.7$.